\documentclass[manuscript, screen, natbib=false, nonacm]{acmart}

\usepackage{capt-of}
\usepackage[table]{colortbl}
\usepackage{algorithm}
\usepackage{algpseudocode}
\usepackage{enumitem}
\usepackage{setspace}

\setlist[itemize]{label=\raisebox{0.35ex}{\scalebox{0.45}{$\blacksquare$}},leftmargin=*,nosep}
\usepackage{graphicx}
\usepackage{siunitx}
\usepackage{wrapfig}

\RequirePackage[
    datamodel=acmdatamodel,
    style=acmnumeric,
    sorting=none
]{biblatex}
\usepackage{xspace}
\usepackage{xcolor}
\newcommand{\rev}[1]{#1}

\newcommand{\ie}{\emph{i.e.}, }
\newcommand{\eg}{\emph{e.g.}, }

\newcommand\blfootnote[1]{%
  \begingroup
  \renewcommand\thefootnote{}\footnote{#1}%
  \addtocounter{footnote}{-1}%
  \endgroup
}

\copyrightyear{2026}
\acmYear{2026}
\setcopyright{none}

\begin{document}

\title[PRESLEY]{Perceptual Refinement of an End-to-End Video Streaming Pipeline via Generative AI Layers}

\author{Emanuele Artioli}
\email{emanuele.artioli@aau.at}
\orcid{0009-0007-9921-0006}
\author{Farzad Tashtarian}
\email{farzad.tashtarian@aau.at}
\orcid{0000-0002-5584-6690}
\author{Christian Timmerer}
\email{christian.timmerer@aau.at}
\orcid{0000-0002-0031-5243}
\affiliation{%
  \institution{Christian Doppler Laboratory ATHENA, Institute of Information Technology (ITEC), Alpen-Adria-Universität Klagenfurt}
  \city{Klagenfurt}
  \country{Austria}
  \postcode{9020}
}

\renewcommand{\shortauthors}{Artioli et al.}

\begin{abstract}
\rev{Traditional codecs treat every region of a frame alike; a generative layer can instead degrade the regions a viewer attends to least and reconstruct them at the client. We present \textbf{PRESLEY}, which extends the prior conference work ELVIS~\cite{elvis} by replacing destructive block removal with adaptive in-place degradation under a removability mask, signaling per-block strength in a bit-packed side channel, and restoring via generative backbones \emph{conditioned} on transmitted visual priors rather than unconditioned in-painting.}
We separate the problem into three goals: choosing which blocks to degrade, degrading them so the encoder spends fewer bits, and restoring them. \rev{Against its predecessor at matched rate, PRESLEY achieves a decisive mean \textbf{\qty{-56.4}{\percent} BD-rate reduction} on delivered background quality across 13 rate ladders spanning multiple codecs and dataset families. Against pristine baselines, PRESLEY defines the operating regime of generative transport: delivering substantial bitrate savings (up to \qty{-29.4}{\percent} BD-rate) and superior background quality (\num{17}/\num{23} sequences) in the target bit-starved regime, while maintaining foreground fidelity bit-exact.}
\rev{We further map where the theoretical headroom in this class of architecture lies. Using an exact leave-one-superblock-out combinatorial oracle as an additive empirical bound, we show that existing complexity heuristics already capture \qty{83.3}{\percent} of bit-cost savings, bounding remaining cost-axis headroom at ${\approx}\qty{5}{\percent}$ of total bitrate. We then identify and model the primary unaddressed axis---post-restoration damage---which disperses widely ($4.9\text{--}8.4$~dB). We prove that this damage is predictable before transmission (held-out $\rho = +\num{0.400}$), establishing the feasibility of transmit-time restorability modeling and defining the roadmap for joint rate-distortion-restoration selection rules.}
\end{abstract}



\maketitle
\blfootnote{Code available at \url{https://github.com/emanuele-artioli/presley}. Extended version of the NOSSDAV~'25 paper ELVIS~\cite{elvis}. Submitted to the ACM TOMM special issue on MMSys and co-located workshops.}

\section{Introduction} \label{sec:introduction}

Video streaming dominates internet traffic, and demand keeps growing with screen resolutions, immersive formats and the number of connected devices.
Traditional video compression methods have achieved remarkable efficiency gains through decades of refinement, with each codec generation from AVC~\cite{overviewAVC} to HEVC~\cite{overviewHEVC} and VP9~\cite{li2019codecs}, to the most recent iterations, VVC~\cite{overviewVVC} and AV1~\cite{overviewAV1} reducing bitrate requirements by approximately 50\% compared to their predecessors~\cite{minopoulos2020codecs-for-live}.
Those gains are increasingly hard-won, and they are spent uniformly: a codec works on blocks of pixels, a level of abstraction at which nothing distinguishes the region a viewer is watching from the region behind it.
Furthermore, treating video frames as collections of pixel blocks gives up the semantic understanding of scene content.
This uniform treatment leads to suboptimal bit allocation: background details consume bits that could be better spent on perceptually important foreground objects, and high-frequency textures are encoded exhaustively even when they could be plausibly reconstructed from context.
Content-aware encoding was proposed for exactly this as early as the 2000s~\cite{1221642, kounoudes2006visual} and has not gained mainstream adoption. 
The obstacles are partly practical (tooling, and semantic analysis inside a real-time encoder), but the binding one is perceptual. Coding neighbouring regions at visibly different quality puts a boundary between them, and a quality discontinuity is more objectionable than a uniformly lower quality at the same rate, so the steps must stay small enough to remain invisible and the achievable saving is capped with them. A generative layer changes that constraint rather than working within it: if the client \emph{reconstructs} the degraded regions instead of displaying them, the delivered frame need not carry the discontinuity that the transmitted one did. We find that it does not. 
Measured on our own output, degraded blocks lying against undegraded ones are damaged \emph{less} than degraded blocks in the interior of a degraded region, by about half a decibel of PSNR, because a block with intact neighbours gives the restorer real detail to work from.
Generative models trained on massive datasets have learned implicit representations of how objects appear and move in video, letting them plausibly reconstruct missing or degraded regions~\cite{propainter, e2fgvi, wang2021realesrgan, instantir}. This opens a paradigm for compression: rather than encoding all visual information, selectively degrade the less critical regions and leverage client-side models to restore them.

\rev{\textbf{PRESLEY} (Perceptual Refinement of an End-to-End Video Streaming Pipeline via Generative AI Layers) realizes it. As an extended version of the NOSSDAV '25 conference paper ELVIS~\cite{elvis}, it advances beyond block removal by degrading adaptively in place, preserving frame structure while reducing quality where it is least perceptible, \ie a generative-AI form of Region-of-Interest encoding, which simplifies codec integration and maintains spatial coherence.}
Posed this way, the problem separates into three goals: \textbf{(1) Selection} --- decide \emph{which} blocks of which frames to degrade. \textbf{(2) Reduction} --- degrade them so that the encoder actually spends fewer bits on them, and relocates those bits to the regions the viewer attends to. \textbf{(3) Restoration} --- recover the degraded regions on the client, as close to the original as a generative prior can bring them.

The key contributions of this work are fourfold:
\begin{itemize}
    \item \textbf{An architecture.} PRESLEY advances the generative streaming paradigm of ELVIS~\cite{elvis} by degrading per block in place under a removability mask, at a strength signalled to the client in a bit-packed side channel, preserving frame geometry while remaining fully compatible with unmodified standard codecs. We evaluate four degradation modalities and price them alongside foreground protection: on content where foreground competes for removability, protecting it preserves perceptual fidelity at a modest \qty{6}{\percent} rate cost.
    \item \textbf{A conditioned restoration path.} We introduce client-side strategies in which a generative model refines each block according to its transmitted strength, and define the pyramid by the model's interface, so the backbone is an interchangeable component. We catalogue ten of them across the transports they are dispatched on.
    \item \textbf{An evaluation protocol.} Fixed-QP rate control, region-restricted perceptual metrics, and paired significance testing with correction over every candidate tried.
    \item \rev{\textbf{A theoretical headroom bound and restorability model.} Moving beyond empirical heuristics, we formulate an exact combinatorial bit oracle that shows existing complexity scores capture \qty{83.3}{\percent} of theoretical bit-cost savings, bounding remaining cost-axis headroom at ${\approx}\qty{5}{\percent}$ of total bitrate under an additive empirical assumption. We then identify and model the second axis---post-restoration damage---proving it is predictable before transmission (held-out $\rho = +\num{0.400}$) from transmit-time spatial features alone, providing the first principled roadmap for rate-distortion-restoration selection.}
\end{itemize}

\section{Background and Related Work} \label{sec:background}

Video streaming relies on maximizing encoding efficiency on the server and adapting to network conditions on the client. With the advent of powerful AI models and capable end devices, it is now possible to share the computational burden, leveraging client-side generative AI to boost delivered video quality.

\subsection{Redundancy Removal}
Traditional codecs --- AVC/H.264~\cite{overviewAVC}, HEVC/H.265~\cite{overviewHEVC}, and the more recent VVC/H.266~\cite{overviewVVC} and AV1~\cite{overviewAV1} --- reduce file size by searching exhaustively for inter-frame motion matches and intra-frame patterns. The approach is highly successful and now facing diminishing returns, further bitrate reduction costing extreme encoding complexity~\cite{codecs-complexity}. More fundamentally, these codecs operate on blocks of pixels, a level of abstraction at which nothing distinguishes one region of a scene from another: every region is treated alike apart from generic heuristics for smooth against detailed areas, which allocates bits without regard to what a viewer is looking at~\cite{8965753}.

\subsubsection*{\textbf{Region-of-Interest coding}}
The idea that not all pixels deserve equal bits is not new, and the standards provide hooks: AVC and HEVC accept per-region QP offsets~\cite{avc_quant_offset}, and a body of work builds rate control on them, detecting important regions with object detectors or saliency models and showing perceptual gains at small bitrate overhead~\cite{roi2006avc, roi2016hevc, roi2014}.
It has nevertheless seen little adoption in general streaming, for four reasons. (1) Tooling: FFmpeg's \texttt{addroi} filter supports only static regions that cannot change per frame~\cite{FFmpegAddRoiFilter}, and the encoder-level interfaces that would allow more are sparsely documented and rarely used~\cite{x264_quant_offsets, x265_cli_qpfile}. (2) Cost: identifying important regions demands semantic analysis inside a codec that must also run in real time. (3) Personalization: what counts as important varies with content and with the viewer, so the driving heuristics are difficult to automate across diverse material~\cite{digitwise}.
The final reason is more subtle, and shapes the design space more than the others. Coding neighbouring regions at visibly different quality places a boundary between them, and a quality discontinuity is more objectionable than a uniformly lower quality at the same bitrate. ROI coding must therefore keep its quality steps small enough to stay invisible, which complicates its logic and caps what it can save. A system that \emph{reconstructs} the degraded regions at the client is not bound by that cap, because the delivered frame need not carry the discontinuity the transmitted one did.

\subsubsection*{\textbf{Semantic Streaming}}
A more radical class of approach transmits high-level descriptors instead of pixels and reconstructs at the client. In video conferencing, SemConf~\cite{semconf} and NVIDIA Maxine~\cite{NVIDIA_Maxine} send sparse facial keypoints and synthesize the talking head at the receiver, achieving very large reductions by transmitting who is moving and how rather than the pixels that show it. The idea is being extended toward general streaming~\cite{jin2025genai_mm_theory}: GenStream~\cite{genstream} transmits skeletal keypoints, camera parameters and a static background model, and reports over \qty{99}{\percent} bandwidth reduction against HEVC on a sports sequence.
These systems show how much redundancy remains once semantics are available, but they are confined to prototypes and favourable scenarios: talking heads are constrained in a way arbitrary content is not, and without standardization the reconstruction method and the bitstream must be designed together, so nothing interoperates with a deployed decoder.
Reducing resolution before transmission and reconstructing it at the client sits between these extremes and is established practice: large services deliver downscaled renditions, and receiver-side super-resolution is studied and deployed to recover them. Two properties bound it. The reduction is uniform over the frame, so no region can be protected; and the client is told only which rendition it received, so its restoration is unconditioned on what was degraded or where.

\subsection{Video Restoration}
The client's task is to reconstruct the decoded video at the highest quality it can, and three families of technique address different parts of it.

\subsubsection*{\textbf{Frame interpolation}} raises frame rate by synthesizing intermediate frames, improving fluidity without raising the transmitted rate~\cite{niklaus2017video}. Modern methods estimate optical flow to do it, adding depth so that closer objects are sampled preferentially through occlusions~\cite{bao2019depth} or self-attention for long-range motion~\cite{shi2022vfit}, but unpredictable motion still produces ghosting where the motion estimate is wrong.

\subsubsection*{\textbf{Super-resolution}} reconstructs high-resolution frames from low-resolution input, from upscaling filters such as bicubic and Lanczos~\cite{resizing_rescaling_opencv} to adversarially trained networks like ESRGAN~\cite{wang2018esrgan}, which optimize for perceptual detail rather than pixel error. Video-specific models add temporal reasoning~\cite{cao2021video}, and interpolation and super-resolution can be combined in one framework~\cite{kim2020fisr}. Cost remains the barrier to live deployment, and because these models operate on low-level features they cannot reason about what an occluded region contains.

\subsubsection*{\textbf{Video in-painting}} reconstructs content that is missing rather than merely degraded. Early methods propagated patches from neighbouring frames~\cite{Patwardhan2005inpainting} and flickered across them; global optimization improved consistency at considerable cost~\cite{Newson2014inpainting}. Generative models changed what is achievable, filling all frames jointly with spatial attention for texture and temporal attention for correspondence~\cite{zeng2020sttn}, or propagating content along estimated motion~\cite{gu2023flow}. What none can do is recover the specific content that was removed, they only synthesize something plausible.

\subsection{Quality Assessment for Generative Video}
Traditional full-reference metrics like PSNR, SSIM, and VMAF~\cite{vmaf} are standard for evaluating codec fidelity. However, they heavily penalize generative reconstructions that are structurally misaligned but perceptually realistic~\cite{onuoha2023field}. 
To accurately assess AI-enhanced video, learned perceptual metrics are required. LPIPS~\cite{lpips} computes similarity in deep feature space, correlating much closer with human judgment for GAN-generated content, as does DISTS~\cite{dists}, which additionally separates structure from texture similarity so that plausible re-textured detail is not penalized as heavily as it would be by pixel-aligned metrics.
Because generative pipelines can also introduce localized hallucinations or temporal jitter, distributional metrics against the reference are often used: the Fr\'echet Inception Distance (FID)~\cite{fid} for spatial realism, and the Fr\'echet Video Motion Distance (FVMD)~\cite{fvmd} for temporal consistency. \rev{The foreground claims in this article rest on LPIPS and DISTS restricted to a mask, because the foreground variants the others expose are bounding-box crops and a box drawn around these sequences contains much of the background the method degrades on purpose. All seven measures are reported at whole-frame scope, where that objection does not apply, in Appendix~\ref{sec:wholeframe}.}

\section{PRESLEY} \label{sec:implementation}

\begin{figure}
    \centering
    \includegraphics[width=0.68\linewidth]{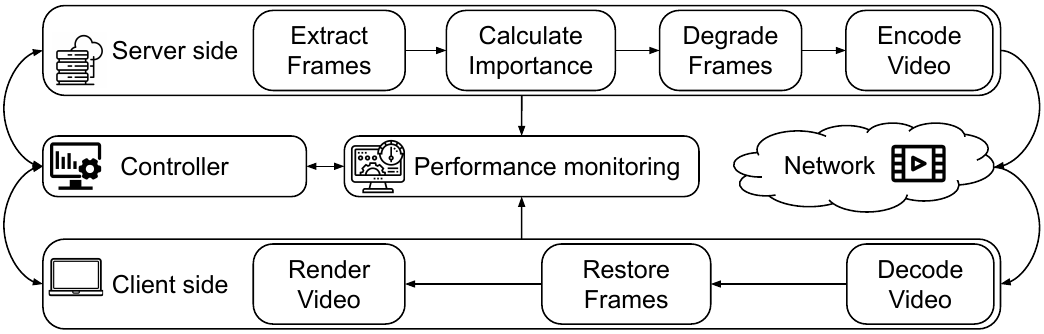}
    \caption{PRESLEY System Architecture Overview.}
    \input{Figures/overview.desc}
\label{fig:overview}
\end{figure}

PRESLEY builds on ELVIS~\cite{elvis}, which integrates a generative layer into a streaming pipeline by \textit{frame shrinking}: the server scores blocks, removes the most removable, and the client stretches frames back to size and in-paints the gaps.
PRESLEY reworks that into a codec-friendly, non-shrinking transport, and supports two degradation families on it. The first keeps every block in place and replaces the content of the most removable ones with a flat fill or with the co-located block from the previous frame: ELVIS's idea of giving a block up entirely, but given up \emph{in place}, so frame geometry never changes and no stretching is needed before restoration. The second punches no holes at all: it \textit{degrades} the most removable blocks in place by lowering their spatial resolution, so every pixel stays partially present in the transmitted stream. Because nothing is discarded, the client can pair it with a \textit{conditioned} restoration model that consumes the degraded pixels directly rather than an in-painter that must hallucinate the region from neighbouring frames.
Both families share the modular framework of Figure~\ref{fig:overview}: server-side preprocessing and removability scoring, selective degradation, standard network transmission, and client-side restoration.
 
\subsection{Server side Degradation Pipeline}
\label{degradation}
The server preprocesses frames --- extraction, rescaling, and lossless encoding into a reference video against which subsequent distortion is assessed --- and then begins the degradation process of Figure~\ref{fig:degrading}. The first step decides which blocks are the best candidates, weighing the trade-off that decision carries: degrading complex blocks lowers the bitrate more than degrading simple ones, but also makes them harder to recreate.

\subsubsection*{\textbf{Removability Calculation.}}

\begin{figure}[h]
    \centering
    \includegraphics[width=1\columnwidth]{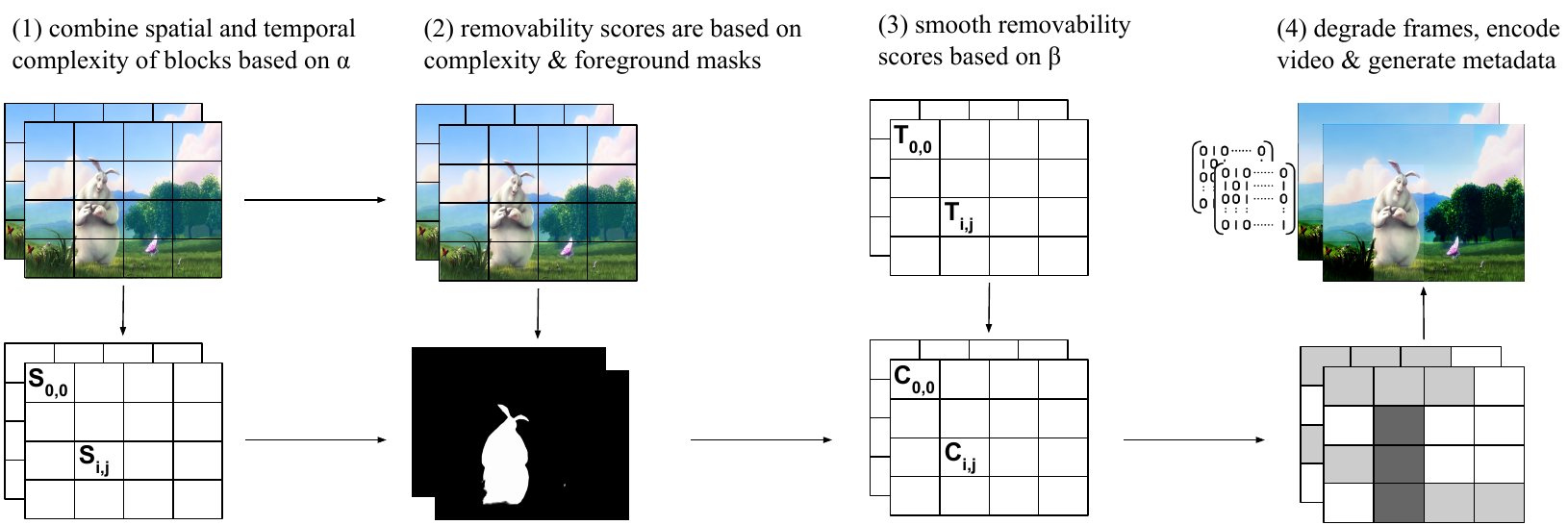}
    \caption{Server side pipeline: removability calculation and adaptive degradation.}
    \input{Figures/degrading.desc}
\label{fig:degrading}
\end{figure}

ELVIS and PRESLEY resolve it by degrading complex blocks and letting the restoration model reconstruct them from spatial and temporal neighbours, prioritizing the background and limiting degradation of the main object, where the viewer's eyes focus.
Equations~\ref{eq:complexity}--\ref{eq:removability} below compute a \emph{removability} score for every block: high where a block is a good candidate for degradation, low where it is not. It requires (1) block-wise spatial ($S$) and temporal ($T$) complexity tensors --- detail within a block, and motion between corresponding blocks in successive frames --- computed at the configured block size with the Enhanced Video Complexity Analyzer (EVCA)~\cite{evca}, and (2) a segmentation tool separating foreground from background. The complexity of block $b_{i,j,n}$ is a weighted aggregation of the two:
\begin{equation}
    C_{i,j,n} = \alpha \cdot S_{i,j,n} + (1 - \alpha) \cdot T_{i,j,n+1},
    \qquad C_{i,j,N-1} = S_{i,j,N-1}
    \label{eq:complexity}
\end{equation}
where $\alpha$ controls the relative influence of spatial and temporal factors (Figure~\ref{fig:degrading}, step (1)). The temporal term is read from frame $n+1$ rather than $n$: EVCA reports the complexity of the transition \emph{into} a frame, so $T_{i,j,n+1}$ is the motion a decision taken at frame $n$ must survive. The final frame has no successor and falls back to spatial complexity alone. Foreground-background segmentation uses object detection models (\eg UFO~\cite{ufo}), returning a mask $M_{i,j,n}$ marking blocks where the object is present ($1$) or absent ($0$); background blocks are given priority by scaling their complexity by a constant $\gamma$ (Figure~\ref{fig:degrading}, step (2)):
\begin{equation}
    \hat{C}_{i,j,n} = \Gamma_{i,j,n} \cdot C_{i,j,n},
    \qquad \Gamma_{i,j,n} = \gamma - (\gamma - 1)\, M_{i,j,n}
   \label{eq:priority}
\end{equation}
The mask $M$ is \num{1} on the foreground and \num{0} on the background, so $\Gamma$ is \num{1} on the foreground and $\gamma$ on the background: a foreground block keeps its complexity unchanged, and a background block has its removability raised by $\gamma$, with $\gamma = \num{10}$ throughout. This is a \emph{soft} priority, shown as step~(2) of Figure~\ref{fig:degrading}: it scales the two populations apart without separating them, so a sufficiently complex foreground block can still outrank a simple background one. Whether that residual overlap matters is a question about content rather than about the score, and we return to it once the selection rule is defined.
To further refine block degradation decisions, the smoothed score $C'_{i,j,n}$ is defined as follows:
\begin{equation}
    C'_{i,j,n} = \beta \cdot \hat{C}_{i,j,n} + (1 - \beta) \cdot \hat{C}_{i,j,n-1},
    \qquad C'_{i,j,0} = \hat{C}_{i,j,0}
    \label{eq:smoothing}
\end{equation}
where $\beta$ is a tunable parameter controlling the degree of temporal smoothing, as shown in Figure~\ref{fig:degrading}, step (3). Values below \num{1} average a block's score with its own value in the previous frame, so a block whose complexity fluctuates frame to frame does not flicker in and out of the selection. Two properties of the form bound what it can express, and both bear on the ablation of Section~\ref{sec:ablation}: smoothing is applied \emph{after} the region priority of Equation~\ref{eq:priority}, against the previous frame's \emph{unsmoothed} score, so a frame's influence decays after exactly one step.
Finally, the smoothed scores are min-max normalized over the whole tensor to give the removability score used by every downstream step:
\begin{equation}
    R_{i,j,n} = \frac{C'_{i,j,n} - \min C'}{\max C' - \min C'} \in [0, 1]
   \label{eq:removability}
\end{equation}
Higher $R$ means more removable. The normalization is what lets a single scalar budget and threshold carry the same meaning across sequences whose EVCA scores differ by orders of magnitude.

This is PRESLEY's answer to the selection goal. Every term entering Equation~\ref{eq:complexity} is a complexity measure, and complexity is a proxy for \emph{the number of bits a block costs the encoder}; the equations that follow reorder that estimate by region, smooth it over time and rescale it. An ideal score would weigh those bits against the other quantity a selection decision depends on --- how faithfully a block can be reconstructed once degraded --- and maximize bits freed per unit of damage surviving restoration. We do not model that second quantity here, for a reason of measurability rather than of principle: it is a property of the restored output, which the server does not have at the moment it must choose, and no transmit-time proxy for it was available when the score was specified. The formulation inherited from ELVIS therefore expresses the numerator alone, and Section~\ref{sec:ablation} takes up what that omission costs.

\subsubsection*{\textbf{Degradation Strategy.}}
Both systems act on the same score. In ELVIS, the blocks with the highest removability are dropped and the frame is shrunk one block-row at a time until the configured shrink amount is reached, so that surviving blocks still repack into a rectangle a codec will accept.
PRESLEY removes nothing. It degrades the most removable regions in place, so frame geometry never changes and the per-row constraint disappears with it.
Block selection is shared across every degradation described below, and turns the score $R$ of Equation~\ref{eq:removability} into a binary map $\sigma$ of degradable blocks in three steps. \emph{A global budget, not a per-row count:} ELVIS removes a fixed number of blocks from every row because its survivors must repack into a rectangle, but PRESLEY keeps frame geometry, so that constraint can be relaxed. PRESLEY takes the globally most-removable $k$ blocks, with $k = \lfloor a \cdot J \rfloor \cdot I$ for a configurable budget $a$ --- the same total the per-row rule would have spent, capped at the number of eligible blocks. \emph{Spatial clustering before ranking:} the score map is smoothed with a small Gaussian kernel before the top-$k$ is taken, so selected blocks form contiguous patches rather than isolated singletons, because a codec's block-level prediction reproduces one large uniform region far more cheaply than the same area of confetti. \emph{Hard foreground exclusion, applied after the blur:} foreground blocks are removed from candidacy outright, so foreground content is never degraded whatever its score.

The exclusion is a design response to what we saw in early runs: on content whose foreground is complex enough to compete, the soft priority of Equation~\ref{eq:priority} left a fraction of foreground blocks inside the selection, and raising $\gamma$ did not clear them, because a factor applied to every background score alike cannot express a constraint that the ranking must never cross a region boundary. What the exclusion buys is a clean attribution rather than a better average. With it on, foreground blocks are carried through from the transmitted frame untouched by the generative layer --- not bit-exact against the source, since they are still lossily coded at the same QP as everything else, but identical across restorers. Any foreground difference between two configurations is therefore the codec's doing and not the restorer's, which is what makes the backbone catalogue of Section~\ref{sec:results} a comparison of backgrounds. The exclusion is applied \emph{after} the clustering blur rather than before, since a blur applied first smears foreground scores across the boundary and lets excluded content back into the ranking through its neighbours.

It is a configuration flag rather than a property of the transport, and turning it off is a coherent operating point: the soft priority of Equation~\ref{eq:priority} still applies, so the foreground is degraded last rather than never, and the mask-restricted metrics measure whatever ends up in the region either way. We enable it throughout this article so that the configuration comparisons isolate one variable at a time, and Section~\ref{sec:exclusion} measures what it costs in bitrate to keep it.

PRESLEY supports two families of degradation modality, differing in how much of ELVIS they carry: \textit{hole-filling} degradations, which take ELVIS's block-removal insight onto the new transport, and \textit{downsampling}, which departs from it entirely.

\textit{Hole-filling degradations.} Selected blocks are replaced in place, either with the flat mean color of the block (\textit{mean\_fill}, referred to as \textit{blackout}) or with the corresponding block copied from the previous degraded frame (\textit{freeze}). Both leave a low-information prior in place of the original content: a flat block costs a codec almost nothing (a single DC coefficient plus strong prediction), and a copied block is typically coded as a near-zero-cost inter-skip. What ELVIS supplies is the idea that the most removable blocks can be given up entirely and synthesized at the client; what is new here is that they are given up \emph{in place}. Keeping frame geometry is what removes the per-row budget, the block shifting and the frame stretching, and it is what makes the global selection described above available to this family at all --- none of which ELVIS's shrink transport can express. These degradations are the transport paired with the video in-painters of Section~\ref{sec:restoration}.

\textit{Downsampling.} Rather than discarding a block's content, downsampling reduces its information density. Each selected block is assigned a strength level $L$, and the transmitted frame is produced by downsampling that block by $2^{L}$ using pixel area relation~\cite{INTER_AREA_OpenCV} and immediately resampling it back to $b \times b$ by bilinear interpolation, so the transmitted frame keeps its original resolution while carrying less high-frequency detail where it was degraded. The level map is
\begin{equation}
    L_{i,j,n} = \sigma_{i,j,n} \cdot \max\big(1,\; \lfloor \lambda \cdot R_{i,j,n} \rceil\big)
   \label{eq:downsampling}
\end{equation}
where $\sigma$ is the block selection just described, $\lambda$ the number of available levels, and $\lfloor \cdot \rceil$ rounds to nearest. Selection and strength are two decisions read off one score, and the equation keeps them consistent: $\sigma$ decides \emph{whether} a block is degraded, $\lambda R$ decides \emph{how far}, and the level is defined on the selected set with a minimum of one, since a selected block at level \num{0} would be a contradiction and would put the realized removal rate below the budget.
The range of $\lambda$ is set by the block size rather than chosen freely. A block of side $b$ at level $L$ is carried at $b/2^{L}$ samples per side, so $\lambda \le \log_2 (b / b_{\min})$. We take $b_{\min} = \num{8}$: below that a block holds fewer samples than the smallest transform a codec applies, and a conditioned restorer has nothing left to condition on --- the degradation has become a hole fill. Graded strength is therefore a large-block instrument: at $b = \num{16}$ the bound admits a single level and Equation~\ref{eq:downsampling} collapses to a fixed $2\times$ reduction on every selected block, while at $b = \num{64}$ it admits three and the deepest QP still arrives as $8 \times 8$.
Four more operators act on the same selection and reduce a block's information without emptying it.

\textit{Blur.} Each selected block is convolved with a Gaussian kernel, one pass per strength level, removing high-frequency detail while leaving low-frequency structure and the block mean in place. It differs from downsampling in what it does to the sampling grid --- downsampling discards samples and reconstructs them by interpolation, blurring keeps every sample and attenuates the detail it carries --- but both leave a prior a conditioned restorer can consume.

\textit{Transform-domain truncation.} Rather than filtering on the pixel plane, this operator writes exact zeros in the codec's own basis. Each selected block is converted to \mbox{YCrCb} and partitioned into $8 \times 8$ sub-blocks --- the smallest transform AV1 applies --- a 2-D DCT is taken per channel, every coefficient outside the top-left $\kappa \times \kappa$ corner is zeroed, and the sub-block is reconstructed by an inverse DCT. The motivation is alignment rather than severity: a pixel-plane low-pass straddles transform boundaries and leaves residual energy the encoder must still code, while zeroing coefficients on the codec's own grid removes exactly the terms it would otherwise spend bits on.

\textit{QP Mapping.} Where codecs support QP maps, per-block offsets can be set proportionally to removability:
\begin{equation}
    QP_{i,j,n} = QP_{base} + (QP_{max} - QP_{base}) \cdot R_{i,j,n}
    \label{eq:qp}
\end{equation}
Here $QP_{base}$ is the codec's baseline QP, $QP_{max}$ the maximum allowable, and $R$ the removability score normalized to $[0,1]$. Two properties distinguish this route. It re-quantizes DCT coefficients on the pixel plane before encoding rather than driving the codec's own per-block control, so the encoder receives an already-quantized frame and quantizes it again; and because available codecs accept only per-GoP QP maps, frame-level maps must be averaged across each GoP, so a block receives the strength its neighbourhood in time received. This route is also where the soft priority of Equation~\ref{eq:priority} does its remaining work: it consumes $R$ directly, with no selection step and so no hard exclusion, which makes $\gamma$ the only foreground protection it has. The same is true of ELVIS's shrink transport. And because $\Gamma$ is applied \emph{before} the temporal smoothing of Equation~\ref{eq:smoothing}, a block whose region membership changes between consecutive frames carries a $\gamma$-scaled value into the smoothed score, so the priority still moves the ranking at foreground boundaries under motion, where membership is least stable.

\textit{Noise Injection.} An alternative degradation reuses the level map of Equation~\ref{eq:downsampling} as rounds of additive Gaussian noise, one per level. It is the one modality that does not reduce a block's information content --- it \emph{adds} high-frequency detail --- so it separates the two things a degradation is asked to do at once: look acceptable to a viewer, and look cheap to a codec.

\subsubsection*{\textbf{Encoding and Metadata.}}
Step~(4) of Figure~\ref{fig:degrading} applies the chosen degradation to the selected blocks, encodes the result with an unmodified codec, and emits the side channel the client needs to undo it. ELVIS shrinks frames during encoding, so its client must know which blocks were removed to restore the original geometry. PRESLEY preserves every block, so each keeps its position while its quality varies, and the client instead requires metadata identifying which blocks were degraded and by how much: the level map $L$ of Equation~\ref{eq:downsampling}, which reduces to a binary mask for the hole-filling degradations and for $\lambda = 1$. The map is bit-plane packed into $\lceil \log_2 (L_{\max} + 1) \rceil$ binary planes, eight blocks to the byte, and DEFLATE-compressed, giving an uncompressed cost of
\begin{equation}
    B_{\text{side}} = \Big\lceil \tfrac{1}{8}\, I \cdot J \cdot N \Big\rceil \cdot \big\lceil \log_2 (L_{\max} + 1) \big\rceil \text{ bytes}
    \label{eq:sidechannel}
\end{equation}
against the $32 \, I J N$ bits an uncompressed integer map would need. This tells the client how many rounds of degradation each block went through, and therefore how many rounds of restoration it needs. The cost is counted inside \texttt{transmitted\_size\_bytes} and therefore inside every rate comparison in this article, and Section~\ref{sec:evaluation} quantifies it as a low single-digit percentage of the transmitted budget.

\subsection{Client side Restoration Pipeline}
\label{sec:restoration}
Restoration is driven by the degradation applied before encoding. In ELVIS, where frames are shrunk, it requires decoding the shrunk file, stretching the frames back to full resolution from the metadata, and passing the video to an in-painting model to generate the missing content.

\begin{figure}[h]
    \centering
    \includegraphics[width=0.8\columnwidth]{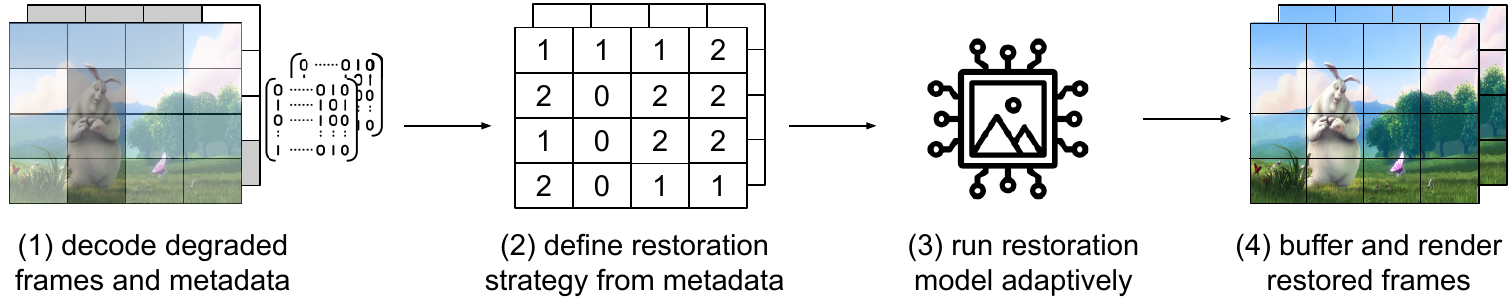}
    \caption{Client side pipeline: metadata parsing, restoration and downstream tasks.}
    \input{Figures/restoration.desc}
\label{fig:restoration}
\end{figure}

Unlike ELVIS, PRESLEY preserves all blocks but adaptively degrades them.
Consequently, restoration for the downsampling family is performed via generative enhancement, tailored to the specific degradation introduced, while restoration for the hole-filling family continues to be performed via video in-painting, exactly as in ELVIS -- the difference being that PRESLEY's transport never altered frame geometry, so no stretching step is needed before in-painting.
\subsubsection*{\textbf{Decoding and metadata parsing.}}
The first step on the client side, as shown in Figure~\ref{fig:restoration}, step (1) is receiving and decoding the degraded video and the metadata.

\subsubsection*{\textbf{Adaptive Restoration.}}
Based on the degradation strategy used at the server, and detailed in the metadata, as per Figure~\ref{fig:restoration}, step (2), PRESLEY applies the corresponding restoration method, orchestrated by Algorithm~\ref{alg:adaptive_restoration}.

\begin{algorithm}[h]
\caption{Adaptive Restoration}
\label{alg:adaptive_restoration}
\begin{algorithmic}[1]
\Require $\mathbf{F}_{in}$: sequence of decoded degraded frames
\Require $\mathbf{L}$: metadata map -- per-block strength levels (Eq.~\ref{eq:downsampling}), or per-block hole flags
\Require $mode \in \{SR, Inpaint\}$: restoration mode, set by the degradation family used at the server
\Ensure $\mathbf{F}_{out}$: restored frame sequence
\If{$mode = SR$}
    \For{each frame $F_{in} \in \mathbf{F}_{in}$ with level map $L$}
        \State $D_{i,j} \gets 2^{L_{i,j}}$
        \Comment{levels are exponents; $D = 1$ means untouched}
        \State $m \gets \max(D)$
        \State $F \gets \text{Resize}(F_{in}, 1/m)$
        \Comment{down to the coarsest block's true resolution}
        \State $s \gets m/2$
        \While{$s \ge 1$}
            \State $F, D \gets \text{Alg.~\ref{alg:sr_round}}(F, F_{in}, D, s)$
            \Comment{Appendix~\ref{sec:restoration-algs}}
            \State $s \gets s/2$
        \EndWhile
        \State Append $F$ to $\mathbf{F}_{out}$
    \EndFor
\ElsIf{$mode = Inpaint$}
    \State $\mathbf{F}_{out} \gets \text{VideoInpaint}(\mathbf{F}_{in}, \text{UpsampleToPixels}(\mathbf{L}))$, composited inside the mask only
\EndIf
\State \Return $\mathbf{F}_{out}$
\end{algorithmic}
\end{algorithm}

If blocks were downsampled, PRESLEY restores them through progressive super-resolution rounds.
The client first downsamples the entire frame to the maximum downsampling factor used. That preliminary reduction is what makes the subsequent upscaling well posed: the received video already matches the reference resolution, so a low-resolution input has to be constructed for the model, and doing it globally and once confines any downsampling artefact to this stage rather than to the high-priority blocks re-injected later.
It then iteratively applies a super-resolution model to upsample the frame by $2\times$ per round. The round is defined by the model's interface, not by a particular model: any restorer taking a frame and returning it at twice the resolution can occupy the slot, which makes the backbone interchangeable.
In each round, as detailed in Algorithm~\ref{alg:sr_round} (Appendix~\ref{sec:restoration-algs}), blocks that have reached their target resolution are replaced with content from the decoded degraded frame resized to the current pyramid level, so only blocks still short of their transmitted detail keep the model's output. That re-injection is the super-resolution family's equivalent of the passthrough compositing described below, with the same consequence: non-degraded blocks are carried through as decoded, so foreground quality does not depend on which backbone is used. The depth of the pyramid follows $\lambda$: a block leaves the loop at the QP matching its own level, so a block degraded once is re-injected earlier, and passes through fewer model applications, than one degraded three times.

If blocks were degraded by a hole fill (blackout/mean\_fill or freeze), the inpainting model receives the decoded frames with the hole mask and synthesizes content for the masked region from neighbouring frames, discarding whatever was transmitted inside the mask. The result is passthrough-composited with the decoded frame --- pixels outside the hole kept exactly as decoded, only those inside taken from the in-painter --- so foreground quality, never degraded by the hard exclusion of Section~\ref{degradation}, is reproduced bit-exact whichever in-painter is used. Because the in-painter discards the transmitted prior rather than consuming it, this path is restricted to the hole-filling degradations; it is the mechanism ELVIS uses, carried over except that no stretching step is required first.
Finally, the restored frames are grouped into segments, buffered and rendered by any traditional video player, or passed to evaluation.
\section{Performance Evaluation}
\label{sec:evaluation}

We evaluate PRESLEY's performance by implementing its pipeline onto a Ubuntu machine, running experiments that link configuration parameter values to variations in output quality, and evaluate results both in terms of quality across several State-of-the-Art (SOTA) image and video metrics, and in terms of throughput added by the degradation/restoration layer. 

\subsection{Evaluation Setup}

Every comparison is against one of three targets: a pristine baseline at the same encoder configuration, ELVIS's block-removal transport at matched rate, or a matched \texttt{none} control running the identical transport with no restorer. Rate is computed as the full transmitted payload, \ie video plus side channel.

\subsubsection*{Metrics, and how a difference becomes a claim}
Foreground and background are scored separately throughout, because the design degrades one and protects the other. \rev{For LPIPS and DISTS the region restriction is a mask weighting of the spatial feature maps rather than a crop, and background PSNR is never a verdict: the flat mean fill scores the best background PSNR of any fill we tested while being the worst-looking. No subjective study was conducted, and claims rest on an exact two-tailed sign test over videos, Holm-corrected across every candidate tried including the losers. Appendix~\ref{sec:metrics} gives both procedures in full.}

\subsubsection*{Restoration backbones}
Which restorer a run dispatches is fixed by its transport. Every model named here was run and is reported. On the \emph{conditioned} path we run Real-ESRGAN~\cite{wang2021realesrgan}, BSRGAN~\cite{bsrgan}, Real-HAT-GAN~\cite{chen2023hat} and Stream-DiffVSR~\cite{shiu2025stream} on the identical pyramid, with Real-ESRGAN carrying the corpus and the others measured against it at a shared operating point; on the \emph{blur} transport, InstantIR~\cite{instantir}, NAFNet~\cite{nafnet} and an \texttt{unsharp} trivial control; on the \emph{hole-filling} path, ProPainter~\cite{propainter} and E2FGVI~\cite{e2fgvi} with Telea's classical in-painting~\cite{telea2004image}. Every configuration is additionally run with the restorer disabled, giving each its own matched unrestored control, which separates the cost of the transport from the recovery the restoration achieves.

\subsubsection*{Corpus and hardware}
Experiments ran on an Ubuntu 22.04 server with an AMD EPYC 7713 (64 cores, \num{256} threads), \qty{1}{TB} of RAM and two NVIDIA RTX A6000 GPUs (\qty{48}{GB} each). Sequences come from DAVIS~\cite{davis}, chosen for variety of composition and camera motion. \rev{DAVIS is small and widely used, so a result measured only on it cannot be separated from the corpus it was tuned against; we therefore add held-out clips from MOSE~\cite{mose,mosev2} and YouTube-VOS~\cite{youtubevos}, two segmentation benchmarks that supply the per-frame object masks this pipeline needs and that neither this work nor ELVIS has ever tuned on. An initial randomized sweep over block size, removal budget, $\alpha$, $\beta$ and three renditions is complemented by the deterministic one-parameter-at-a-time ablation of Section~\ref{sec:ablation}.}

\subsection{Hole-filling Evaluation}
\label{sec:hole-filling transport-eval}

Table~\ref{tab:bdrate} evaluates bitrate reduction under x265 across four fixed QPs (30/32/34/37) at \num{640}$\times$\num{360}. On sequences such as \texttt{bear} and \texttt{camel}, the hole-filling transport achieves substantial bitrate savings of $-29.4\%$ and $-20.6\%$ BD-rate alongside foreground gains of $+1.02$~dB and $+0.62$~dB BD-PSNR, demonstrating effective bit relocation from background to foreground. On other sequences such as \texttt{dog}, \texttt{pigs}, and \texttt{india}, the codec requires more bits at these settings. Across the wider evaluation that follows, we analyze how this trade-off behaves across codecs, dataset families, and operating regimes.

\begin{table*}[t]
\centering
\begin{minipage}[t]{0.46\textwidth}\centering
\small
\begin{tabular}{lrrr}
\toprule
Video & FG BD-PSNR (dB) & FG BD-rate (\%) & Overlap \\
\midrule
bear   & $+1.02$ & $-29.4$ & 0.59 \\
camel  & $+0.62$ & $-20.6$ & 0.70 \\
tennis & $+0.16$ & $-3.8$  & 0.74 \\
dog    & $-0.34$ & $+13.8$ & 0.70 \\
pigs   & $-0.50$ & $+21.2$ & 0.69 \\
india  & $-0.40$ & $+10.6$ & 0.78 \\
\bottomrule
\end{tabular}
\caption{Hole-filling (blackout fill, block size~8, ProPainter, passthrough compositing, \texttt{fg\_protect}) vs.\ the plain x265 baseline: foreground BD-PSNR and BD-rate over four matched fixed-QP points (30/32/34/37) per video at \num{640}$\times$\num{360}. Negative BD-rate indicates fewer bits for equal quality. ``Overlap'' is the shared log-bitrate fraction.}
\label{tab:bdrate}
\end{minipage}\hfill
\begin{minipage}[t]{0.50\textwidth}\centering
\small
\setlength{\tabcolsep}{3.5pt}
\begin{tabular}{@{}lrrrr@{}}
\toprule
 & \multicolumn{1}{c}{\textit{best}} & & & \multicolumn{1}{c}{\textit{worst}} \\
 & QP\,32 & QP\,37 & QP\,42 & QP\,47 \\
\midrule
Clips saving bits    & \num{19}/\num{23} & \num{16}/\num{23} & \num{5}/\num{23} & \num{1}/\num{23} \\
Median rate change   & $-6.6\%$ & $-2.5\%$ & $+10.4\%$ & $+31.4\%$ \\
\midrule
Median BG gap        & $+0.0409$ & $+0.0286$ & $+0.0131$ & $-0.0153$ \\
Clips with better BG & \num{0}/\num{23} & \num{2}/\num{23} & \num{3}/\num{23} & \num{17}/\num{23} \\
\midrule
Median FG gap        & $+0.0047$ & $+0.0028$ & $+0.0044$ & $+0.0007$ \\
\bottomrule
\end{tabular}
\caption{The rate--quality trade across the QP sweep, PRESLEY against the pristine baseline at matched QP on \num{23} clips. Gaps are LPIPS differences (PRESLEY minus baseline, negative favors PRESLEY). Foreground (FG) differences remain well within the \num{0.05} perceptual margin at all QPs; the background (BG) reflects the trade-off.}
\label{tab:trade}
\end{minipage}
\end{table*}

\begin{figure*}[t]
    \centering
    \begin{minipage}[b]{0.53\textwidth}\centering
        \includegraphics[width=\linewidth]{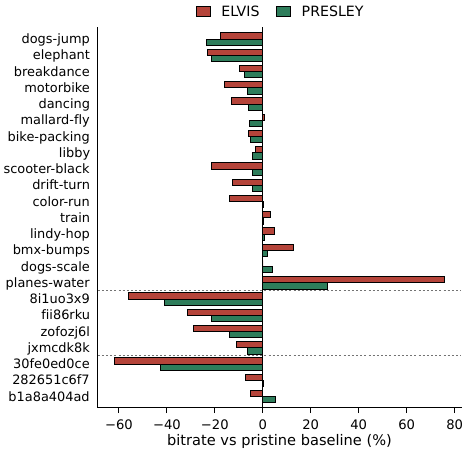}
        \caption{Bitrate delta against the pristine baseline on \num{23} clips where baseline, ELVIS, and PRESLEY all exist at both QPs, grouped by dataset family. Left of zero indicates bitrate savings.}
        \input{Figures/breadth.desc}
        \label{fig:breadth}
    \end{minipage}\hfill
    \begin{minipage}[b]{0.44\textwidth}\centering
        \includegraphics[width=\linewidth]{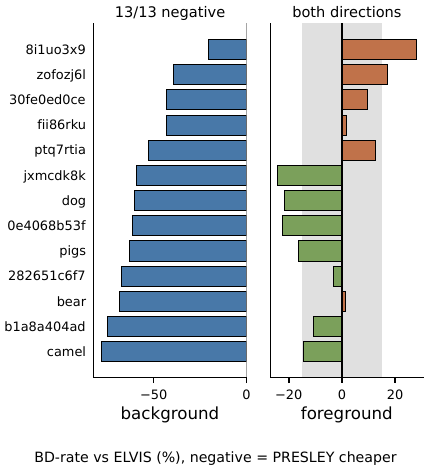}
        \caption{PRESLEY against ELVIS at matched rate on \num{13} rate ladders across two codecs and three dataset families. \emph{Left:} delivered background LPIPS after restoration, consistently favoring PRESLEY on all 13 ladders. \emph{Right:} foreground LPIPS.}
        \input{Figures/ratematched.desc}
        \label{fig:ratematched}
    \end{minipage}
\end{figure*}

\rev{Figure~\ref{fig:breadth} broadens this evaluation across 23 clips from DAVIS, MOSEv2, and YouTube-VOS where the pristine baseline, ELVIS and PRESLEY all exist at both QPs. ELVIS saves bits on \num{17} of \num{23} clips (median \qty{-10.9}{\percent}) and PRESLEY on \num{15} of \num{23} (median \qty{-5.0}{\percent}), and on both \textbf{the sign splits \emph{within} dataset families rather than between them}.} Foreground protection survives outside DAVIS on both, but on different metrics: PRESLEY's foreground LPIPS delta clears the margin at none of \num{47} operating points, while ELVIS's exceeds it at \num{44} of \num{50} even though its foreground \emph{PSNR} moves \num{-0.17}~dB against PRESLEY's \num{-0.16}. A learned metric compares deep features whose receptive fields reach beyond the block, so an emptied neighbour changes a protected block's features without changing its pixels. We therefore report foreground protection on PSNR wherever the transport leaves holes, and on LPIPS where it does not.

The two QPs of Figure~\ref{fig:breadth} sit at the comfortable end of the range. Table~\ref{tab:trade} takes the same configuration across a four-QP sweep on \num{23} clips, illustrating how generative streaming maps onto the operating range. At comfortable bitrates (QP\,\num{32}), the transport frees bits on \num{19} of \num{23} sequences at a median \qty{-6.6}{\percent}, while the baseline retains near-transparent background fidelity. In the target bit-starved regime (QP\,\num{47}), where standard codecs are severely quality-limited, PRESLEY's generative restoration delivers a superior background on \num{17} of \num{23} sequences (median LPIPS gap $\num{-0.0153}$), while holding foreground fidelity strictly within the perceptual margin. This confirms that generative layers provide their greatest perceptual payoff precisely where conventional codecs starve.

QP-matching therefore cannot tell the whole story. At matched \emph{QP}, PRESLEY spends \qty{23.1}{\percent} and \qty{14.9}{\percent} more bits than ELVIS at QP\,32 and \,37, because it transmits every block where ELVIS blanks a quarter of them; hence, the fundamental comparison between the two generative paradigms is evaluated at matched \emph{rate} on delivered quality. \rev{There, Figure~\ref{fig:ratematched} demonstrates that \textbf{PRESLEY comprehensively outperforms ELVIS on delivered background quality across all \num{13} rate ladders, achieving a decisive mean BD-rate reduction of \qty{-56.4}{\percent}} ($p=\num{0.000244}$, strictly clearing family-wise Holm correction across all \num{204} project configurations).} By preserving a degraded visual prior and restoring it with conditioned super-resolution rather than discarding content and relying on unconditioned in-painting, PRESLEY recovers high visual fidelity at a fraction of the rate required by prior generative transports. Furthermore, foreground quality shifts in both directions across ladders without systematic penalty, confirming that background gains carry no hidden foreground cost.

\subsubsection*{Operating point stability}
\label{sec:regime-stability}
A generative transport might only pay off in a favourable corner of the rate range. With the ladder as the unit and the median over the two lowest-quality QPs minus the median over the two highest as the statistic, the contrast is negative on \num{5} of \num{13} at an exact two-tailed $p=\num{0.58}$ --- which alone establishes nothing, since a difference test that fails to reject supports no conclusion about the size of the effect. We therefore tested the proposition directly, as an equivalence: \textbf{the mean contrast is demonstrably smaller than the reporting margin}, with a \qty{90}{\percent} bootstrap interval of $[\num{-0.017}, \num{+0.013}]$ background LPIPS, four times tighter than the \num{0.05} we report against. Equivalence of the \emph{mean} is not equivalence of every ladder: two of thirteen shift by more (\texttt{camel} at $\num{-0.0585}$, a held-out YouTube-VOS clip at $\num{-0.0535}$), both with PRESLEY gaining as the QP starves. \emph{Restoration} holds across the operating range; \emph{reduction} is regime- and content-dependent. Appendix~\ref{sec:robustness} bounds this further with a detector-derived mask, and with \num{1280}$\times$\num{720} and \num{1920}$\times$\num{1080} ladders.

\subsubsection*{Comparison with standard codec ROI and neural codecs}
Standard codec tools offer partial solutions. \rev{As detailed in Appendix~\ref{sec:roi-detail}, Kvazaar's native ROI encoding relocates bits to improve foreground quality by a median $+\num{0.51}$~dB PSNR across \num{16} of \num{17} points at unchanged bitrate, but cannot restore the sacrificed background ($\num{-0.54}$~dB PSNR loss on 8 of 8 points). SVT-AV1's ROI shows no measurable effect.} That relocation is not free: the encoder buys foreground fidelity by spending less elsewhere, and an x265 adaptive-quantization control is indistinguishable from its baseline. PRESLEY instead pairs targeted degradation with generative client-side reconstruction. The stages cannot be stacked, for a structural reason rather than a sampling gap --- the only encoder exposing a usable ROI facility is not the one the hole-filling results run through --- and compared separately they do different things rather than more of one thing: codec ROI is the only stage that improves the foreground, while the degrade-and-restore stages trade background quality for bits. Appendix~\ref{sec:extra-tables} sets the four stages side by side. \rev{Finally, evaluating the neural codec baseline HNeRV (Appendix~\ref{sec:retired}) shows that once serialized model weights are accounted for in the transmitted payload, it requires \numrange{22.7}{178.1}$\times$ more bitrate than x265 at matched foreground quality, confirming that standard hybrid codecs combined with lightweight generative layers remain far more practical.}

\subsection{Restoration and Regime Results}
\label{sec:results}
\begin{figure}[t]
\centering
\includegraphics[width=\textwidth]{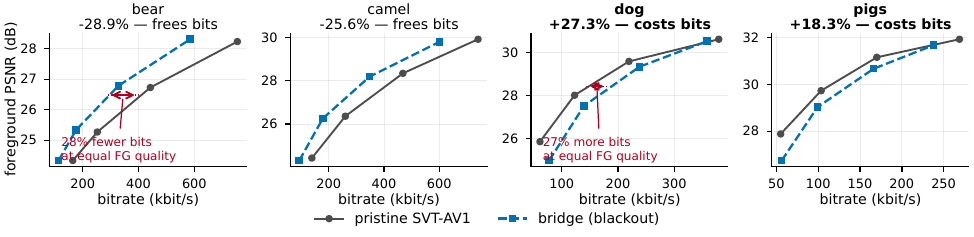}
\caption{The starved-regime result as rate--distortion ladders rather than Bjøntegaard scalars. Each panel is one video over four recalibrated fixed QPs on SVT-AV1, the hole-filling configuration being \texttt{blackout} at block size~16. The claim is horizontal: a curve \emph{left} of the baseline delivers the same foreground quality for fewer bits, and on \texttt{dog} and \texttt{pigs} (bold) it lies right instead.}
\input{Figures/regime_reversal.desc}
\label{fig:regime-reversal}
\end{figure}

Table~\ref{tab:bdrate} measured reduction under x265. \rev{We repeated it under SVT-AV1 across the operating range, holding the transport fixed (block size~16, ProPainter, \texttt{fg\_protect}) and moving only the QP, on four videos spanning the corpus's range of foreground area and camera motion. Figure~\ref{fig:regime-reversal} gives the four rate--distortion curves.
\textbf{Reduction pays where the encoder is starved.} On \texttt{bear} and \texttt{camel} the transport frees \qtyrange{25.6}{28.9}{\percent} of bits at equal foreground PSNR, every point-wise delta under \qty{0.5}{dB}; at comfortable bitrates the same configuration costs \qty{16.7}{\percent} and \qty{40.8}{\percent} more, so the saving belongs to the starved end of the range rather than to the method in general. On \texttt{dog} and \texttt{pigs} it costs bits instead, and the choice of fill does not rescue them. That two-two split is not the corpus rate: across the \num{23} clips of Figure~\ref{fig:breadth} the hole-filling transport saves bits on \num{17} and the conditioned one on \num{15}. These four are the clips we can show as full curves, chosen to include both outcomes.
Two independent mechanisms agree, which makes it a content result rather than a component bug: the same reversal appears in the conditioned configuration of Figure~\ref{fig:breadth}, on the same two videos, through a different degradation and a different restorer. We looked for a content attribute that predicts the flip and did not find one --- foreground area is the obvious candidate and is refuted, \texttt{dog} sitting at \num{0.114} against \texttt{bear}'s \num{0.111} and reversing hardest. We report the non-replication rather than the average over it, and regard identifying that attribute as the most useful open problem this article leaves.}

The curves show what a Bj\o{}ntegaard scalar cannot: on \texttt{dog} and \texttt{pigs} they converge as the QP gets harsher, so the advantage reverses within the sweep rather than merely being absent. Table~\ref{tab:trade} identifies the mechanism --- the transport exchanges rate for background quality at a rate that itself changes with the QP, so a scalar integrated across that crossover sums a cheaper-and-poorer region with a dearer-and-better one and reports them as one effect. Two studies of the same configuration can therefore disagree purely on the range they integrate over.

\begin{figure*}[t]
    \centering
    \includegraphics[width=\textwidth]{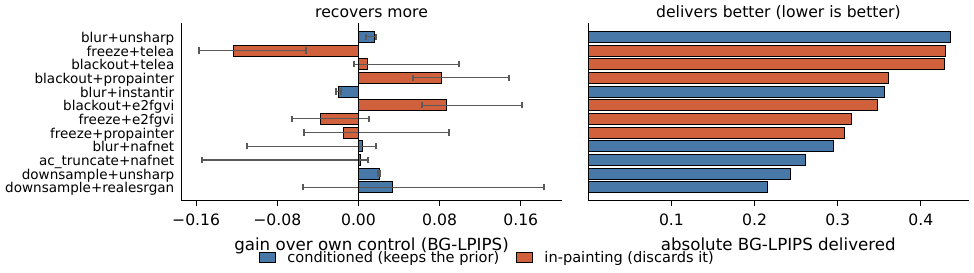}
    \caption{Every restorer we ran, on the transport it is dispatched on, each against its
    own exactly matched unrestored control. \emph{Left:} gain over that control. \emph{Right:}
    the absolute background LPIPS the viewer receives. The panels disagree, because gain is
    bounded by how much each transport destroyed.}
    \input{Figures/restorer_catalogue.desc}
    \label{fig:restorers}
\end{figure*}

\subsubsection*{Does client-side restoration bring the degraded background back toward the original?} Figure~\ref{fig:restorers} answers it for every backbone we ran, each against its own exactly matched unrestored control, differing only in whether a restorer ran. The two panels disagree. Ranked by \emph{gain} over its control, hole-filling wins; on the \emph{absolute} quality a viewer receives the ordering inverts. Each configuration's gain is bounded by how much its own transport destroyed, so a large gain only reports an ambitious transport, and the choice between families is settled on the absolute panel: \textbf{the transport that keeps a prior wins on delivered quality even though it leaves its restorer less to do.}

\begin{figure*}[t]
    \centering
    \includegraphics[width=1\textwidth]{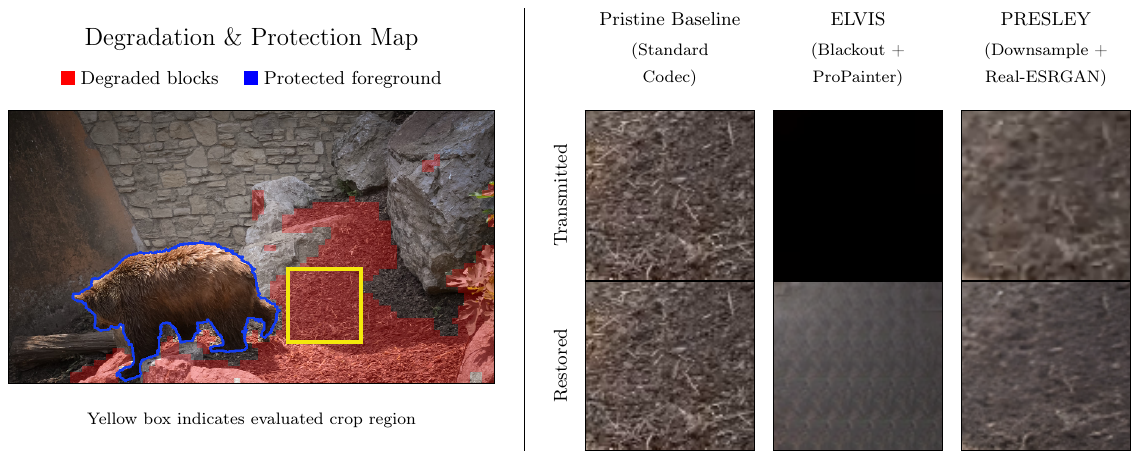}
    \caption{Spatial selection, transport, and restoration visual comparison. \emph{Left:} Selected degradable blocks shaded in red, protected foreground outlined in blue, with the yellow bounding box indicating the evaluated crop. \emph{Right:} Visual patch comparison across pipeline configurations (Pristine baseline under standard lossy codec, ELVIS blackout with ProPainter inpainting, and PRESLEY adaptive downsampling with Real-ESRGAN super-resolution), showing the transmitted representation on the wire (top row) and the restored output delivered to the client (bottom row).}
    \input{Figures/qualitative.desc}
    \label{fig:qualitative}
\end{figure*}

Figure~\ref{fig:qualitative} illustrates the full pipeline on a representative frame: the selector places degradation exclusively on the background while preserving foreground integrity, and the comparative crops show what each configuration puts on the wire (top) versus what its restorer recovers (bottom). While ELVIS transmits empty blocks and in-paints them from neighbouring frames, PRESLEY transmits a low-resolution prior that its restorer sharpens. The restored crop is sharp again, though fine high-frequency texture is plausibly synthesized rather than recovered bit-exact. The mechanism is architectural: models dispatched on hole transports are in-painters (e.g., ProPainter's inference masks out the transmitted prior by construction), whereas conditioned restorers consume those transmitted pixels directly, separating the two families. Freeze proves it from the other side: an in-painter handed a frozen block discards a prior already close to correct, and every freeze pairing sits at or below zero gain.

To check the ordering is not an artifact of the mask, we repeated the comparison on whole-frame metrics, which carry no region restriction. Paired within an operating point, the conditioned configuration beats ELVIS on \num{121} of \num{121} ladder points on whole-frame LPIPS and \num{36} of \num{36} on VMAF, by \num{11.4}~dB of PSNR and \num{25.5} VMAF points at the median. On PSNR and LPIPS it beats blackout in-painting on \num{12} of \num{12} shared operating points ($p=\num{0.0005}$) and freeze in-painting on \num{13} of \num{14}, so it leads every hole-filling configuration. Every measure Section~\ref{sec:background} surveys ranks the configurations as the masked metrics do; Appendix~\ref{sec:wholeframe} gives all seven.

\textbf{No backbone separates from Real-ESRGAN.} The right-hand panel of Figure~\ref{fig:restorers} places every conditioned alternative --- BSRGAN, Real-HAT-GAN~\cite{chen2023hat} and Stream-DiffVSR~\cite{shiu2025stream} --- within a band narrower than the incumbent's perceptual margin, on the same transport and operating point, with Stream-DiffVSR costing roughly twice as much to run. At two paired videos none can reach significance under our own rule, so we report them as a catalogue that failed to separate rather than as demonstrated ties: the evidence supports keeping Real-ESRGAN on throughput, not a claim that it is best. On the blur transport InstantIR does not beat a trivial \texttt{unsharp} control costing ${\approx}12$\,s against ${\approx}774$--$837$\,s, and neither it nor NAFNet delivers meaningful gain against the transmitted video, so that transport's restorer question is open. Both need float32 --- NAFNet overflows LayerNorm and SCA in fp16, Real-HAT-GAN's Softmax overflows to NaN --- so half precision is not safe to assume.

Finally, the part that separates the two goals: \texttt{dog} and \texttt{pigs} produce the largest restoration gains anywhere in this article while being exactly the videos on which reduction fails outright. Restoration quality and reduction are not one axis.

\begin{table}[t]
\centering
\caption{The removal budget below and above its threshold, on the two videos where it was swept: SVT-AV1 at fixed QP, \num{640}$\times$\num{360}, block size~8, downsampling with Real-ESRGAN. Below a quarter of the blocks the budget buys almost no rate while background quality falls steadily, so \num{0.25} is the smallest budget that pays for the damage it does.}
\label{tab:budget}
\small
\begin{tabular}{lrrrr}
\toprule
Removal budget & \num{0.10} & \num{0.25} & \num{0.50} & \num{0.75} \\
\midrule
\texttt{bear} bitrate (kbit/s)  & \num{607.0} & \num{614.2} & \num{451.4} & \num{415.4} \\
\texttt{bear} background LPIPS  & \num{0.141} & \num{0.193} & \num{0.246} & \num{0.291} \\
\midrule
\texttt{camel} bitrate (kbit/s) & \num{694.9} & \num{693.3} & \num{570.5} & \num{531.9} \\
\texttt{camel} background LPIPS & \num{0.110} & \num{0.123} & \num{0.168} & \num{0.195} \\
\bottomrule
\end{tabular}
\end{table}

Two further properties are worth recording. For the winning conditioned pipeline over four starved fixed-QP points per video, reduction is a curve-level win at \qtyrange{13.8}{16.1}{\percent} fewer bits for equal foreground PSNR. And the removal budget has a threshold below which it does nothing: swept at the lowest-quality QP it moves bitrate by under \qty{1.2}{\percent} between \num{0.10} and \num{0.25}, then by \qty{-22.1}{\percent} between \num{0.25} and \num{0.50}, while background damage rises steadily throughout (Table~\ref{tab:budget}). Degrading fewer than a quarter of the blocks therefore costs quality without buying rate, which is why \num{0.25} is the budget used throughout.

\subsection{Conditioned Restoration Wins on Quality; Bitrate and Cost Order the Arms Differently}
\label{sec:frontier}

\begin{figure*}[t]
\centering
\begin{minipage}[b]{0.53\textwidth}\centering
\small
\setlength{\tabcolsep}{3.5pt}
\begin{tabular}{llrrrl}
\toprule
Candidate & Objective & $n$ & Base wins & $p_{\text{Holm}}$ & Sig.? \\
\midrule
\texttt{blur+nafnet}         & quality & 11 & 11/11 & 0.014 & yes \\
\texttt{blackout+propainter} & quality & 10 & 10/10 & 0.025 & yes \\
\texttt{freeze+propainter}   & quality & 10 & 10/10 & 0.025 & yes \\
\texttt{ac\_truncate+nafnet} & bitrate & 10 & 10/10 & 0.027 & yes \\
\midrule
\texttt{ac\_truncate+nafnet} & quality & 10 & 9/10  & 0.236 & no \\
\texttt{blackout+propainter} & bitrate & 10 & 1/10  & 0.279 & no \\
\texttt{freeze+propainter}   & bitrate & 10 & 9/10  & 0.279 & no \\
\texttt{blur+nafnet}         & bitrate & 11 & 5/11  & 1.000 & no \\
\bottomrule
\end{tabular}
\captionof{table}{Every configuration comparable to \texttt{downsample+realesrgan} at our significance floor, paired within a fixed QP. ``Base wins'' counts videos on which \texttt{downsample+realesrgan} is ahead; $p_{\text{Holm}}$ is corrected within a family of 14 candidates per objective. Quality is background LPIPS, bitrate the actual transmitted rate against the matched pristine baseline. Rows marked \emph{no} are comparisons that were run and did not separate at this sample size, reported so the comparison is not selected on its outcome; they are not claims of equality. The two objectives disagree.}
\label{tab:frontier}
\end{minipage}\hfill
\begin{minipage}[b]{0.44\textwidth}\centering
\includegraphics[width=\linewidth]{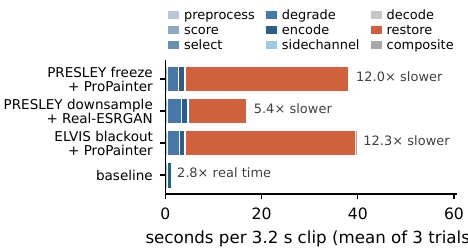}
\caption{Cost per pipeline stage, three trials per cell on a pinned GPU, cells
    mixing device populations refused rather than averaged. Restoration dominates every
    configuration that has one; block selection costs about a thousandth of the pipeline.}
\input{Figures/stage_timing.desc}
\label{fig:stage-timing}
\end{minipage}
\end{figure*}

The comparisons so far run within a family. Ranked that way Real-ESRGAN carries the conditioned path, ProPainter the hole-filling path and NAFNet the blur path, none separating from its nearest rival by more than the perceptual margin. Table~\ref{tab:frontier} sets those family winners against each other, which is the comparison a deployer faces --- and it faces three axes at once, delivered background quality, transmitted bits, and the cost of restoring, which order the configurations differently, so none is simultaneously best. The first is this article's clearest configuration-level result: \textbf{degrading by downsampling and restoring with a super-resolver conditioned on what was actually transmitted delivers the best background quality of any configuration we tested}. Table~\ref{tab:frontier} reports every configuration comparable to it at our significance floor.

On background quality the ordering is settled: the conditioned configuration beats all three in-painting and deblurring configurations on every video tested. On bitrate it inverts --- \texttt{ac\_truncate+nafnet} loses the rate axis on all ten videos ($p=\num{0.0020}$ uncorrected), our one significant bitrate comparison, and it is the configuration the baseline beats on quality. \textbf{Winning quality and winning rate are not the same configuration.} 

Figure~\ref{fig:stage-timing} measures execution time per stage on three \num{640}$\times$\num{360} clips (\numrange{60}{90} frames), three trials per cell on a pinned GPU. Restoration dominates by enough to locate the engineering problem precisely: the conditioned configuration runs \num{5.4}$\times$ \emph{slower} than real time against \num{12.0}$\times$ for the in-painting ones, so the configuration that wins on quality is also the closest to real time, and closing that gap is a restorer-side problem. Everything the server does is comfortably real time --- degradation costs more than encoding, so the bottleneck there is our own per-block loop rather than the codec, and block selection is under \qty{0.1}{\percent} of the pipeline. \textbf{These are \num{360}p figures and the factor scales with pixel count, so we scope the system to offline transcoding and edge-cache pre-population until a restorer closes the remaining \num{5.4}$\times$.}

\subsection{Block Selection and Headroom Analysis}
\label{sec:ablation}

We analyze the block selection mechanism along three dimensions: the efficiency of the cost model, parameter sensitivity, and predictive modeling of post-restoration damage.

\subsubsection*{Placement and foreground exclusion}
\label{sec:exclusion}
Block selection evaluates two distinct mechanisms: ranking blocks by removability score versus random placement, and hard foreground exclusion versus soft priority weighting.

\begin{figure*}[t]
\centering
\begin{minipage}[b]{0.48\textwidth}\centering
\small
\begin{tabular}{@{}lrr@{}}
\toprule
Video & Score ranking & Random map \\
\midrule
\texttt{bear}         & $+80.9\%$  & $+154.4\%$ \\
\texttt{camel}        & $+27.5\%$  & $+121.6\%$ \\
\texttt{dogs-jump}    & $-1.6\%$   & $+79.1\%$  \\
\texttt{pigs}         & $+23.6\%$  & $+75.7\%$  \\
\texttt{dog}          & $+13.2\%$  & $+27.2\%$  \\
\texttt{bike-packing} & $+219.6\%$ & $+194.9\%$ \\
\bottomrule
\end{tabular}
\captionof{table}{Removability score ranking vs.\ seeded random block selection at matched budget, strength, clustering, and foreground exclusion. Values are BD-rates on background LPIPS against each video's pristine baseline (lower is better). The score wins on 5 of 6 sequences by a median of \num{52.1} percentage points.}
\label{tab:placement}
\end{minipage}\hfill
\begin{minipage}[b]{0.48\textwidth}\centering
\includegraphics[width=\linewidth]{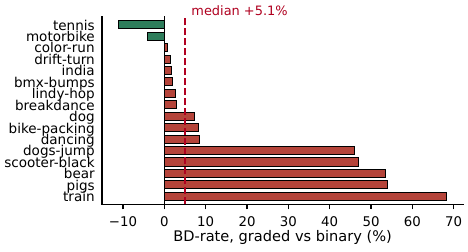}
\caption{Graded multi-level degradation vs.\ uniform binary degradation per sequence at $b=\num{64}$ and \num{1920}$\times$\num{1080}. Positive BD-rate indicates graded requires more bits for equal background quality (losing on 14 of 16 sequences, $p=\num{0.0042}$).}
\input{Figures/graded.desc.tex}
\label{fig:graded}
\end{minipage}
\end{figure*}

As shown in Table~\ref{tab:placement}, ranking blocks by the removability score outperforms a smoothed random selection baseline on \num{5} of \num{6} sequences, providing a median BD-rate improvement of \num{52.1} percentage points on background LPIPS.

\textbf{The exclusion is worth about half a decibel.} On six sequences chosen because a purely score-based selection reaches their foreground at all, dropping it costs foreground PSNR on \num{6} of \num{6}, a median \num{0.56}~dB at $p=\num{0.031}$, while saving \qty{6.0}{\percent} of bitrate; on foreground LPIPS it moves \num{0.0013} and clears no margin anywhere. That is a saving this article declines to take, and the exclusion is a default rather than a property of the transport --- the soft priority alone already confines damage to the least important foreground blocks. Two bounds: the six sequences were chosen \emph{because} selection reaches their foreground, so \qty{6}{\percent} is what the exclusion costs where it binds, not across the corpus; and at six sequences the exact test floors at \num{0.031}, the smallest suite that can be significant at all. What it buys instead is attribution --- with the foreground untouched by the generative layer, any foreground difference between configurations is the codec's doing, which is what lets the backbone catalogue above be read as a comparison of backgrounds.

\subsubsection*{Cost axis headroom bound}
\begin{figure*}[t]
    \centering
    \begin{minipage}[b]{0.48\textwidth}\centering
        \includegraphics[width=\linewidth]{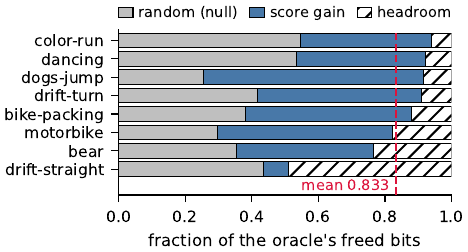}
    \end{minipage}\hfill
    \begin{minipage}[b]{0.48\textwidth}\centering
        \includegraphics[width=\linewidth]{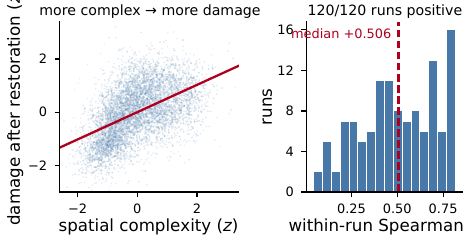}
    \end{minipage}
    \caption{The two axes of the selection objective. \emph{Left:} removability score vs.\ an exact leave-one-superblock-out bit oracle (hatched area indicates remaining headroom). \emph{Right:} post-restoration damage vs.\ spatial complexity, standardized per run.}
    \input{Figures/selection_cost.desc}
    \input{Figures/restorability.desc}
    \label{fig:selection-axes}
\end{figure*}

\rev{The removability score estimates the bitrate contribution of individual blocks. To quantify how closely this heuristic approximates optimal bit allocation, the left panel of Figure~\ref{fig:selection-axes} evaluates the score against an exact leave-one-superblock-out combinatorial oracle across eight probe sequences. Selecting the top quarter of blocks by removability score captures a mean of \num{83.3}\% of the bits identified by the exact oracle. Because the oracle's top quarter accounts for \qty{30.1}{\percent} of total frame bits, \textbf{the remaining headroom on the bit-cost axis is at most ${\approx}\qty{5}{\percent}$ of total bitrate}, measured against the random-selection null the figure draws as the base of every bar. This bound evaluates superblocks independently under the leave-one-out probe, serving as an additive empirical upper bound on cost-model efficiency given inter-block prediction dependencies in hybrid video codecs. A better cost model is therefore not where major gains lie.} One exception: on \texttt{drift-straight} the score barely clears that clip's own null, with a rank correlation whose bootstrap interval contains zero.

\subsubsection*{Parameter sensitivity and degradation levels}
\begin{table}[t]
\centering
\caption{Parameter sensitivity: median within-group spread, a group fixing every other setting and varying only the named parameter. The removal budget is the only lever on either axis. \textbf{Descriptive only} --- the number of videos is below the threshold this article requires for a significance claim. $\alpha$ and $\beta$ are absent because they were measured on foreground PSNR, and putting the two quantities on one scale would compare unlike things.}
\label{tab:ablation}
\small
\begin{tabular}{lrrr}
\toprule
Parameter & $\Delta$bitrate (pp) & $\Delta$BG-LPIPS & Videos \\
\midrule
Removal budget                  & \num{26.38} & \num{0.1508} & \num{2} \\
Block size                      & \num{9.08}  & \num{0.0151} & \num{4} \\
Foreground protection           & \num{5.47}  & \num{0.0024} & \num{2} \\
Mask noise (dilate/erode)       & \num{4.91}  & \num{0.0083} & \num{2} \\
Mask source (GT/YOLO)           & \num{4.20}  & \num{0.0038} & \num{7} \\
Selection rule                  & \num{1.20}  & \num{0.0031} & \num{6} \\
Graded levels                   & \num{1.07}  & \num{0.0149} & \num{8} \\
Graded levels, oracle-assigned  & \num{0.56}  & \num{0.0050} & \num{8} \\
\bottomrule
\end{tabular}
\end{table}

\rev{Table~\ref{tab:ablation} summarizes sensitivity to individual knobs. The removal budget is the primary lever at \num{26.38} percentage points of bitrate, and everything else moves it by under a third as much; swept over $\{0, 0.25, 0.5, 0.75, 1.0\}$, $\alpha$ and $\beta$ move foreground PSNR by \qtyrange{0.03}{0.05}{dB} across their entire range. Both comparisons are \emph{descriptive} --- the number of videos is below what this article requires for a significance claim --- but the magnitudes are not close. That inertness is a deployment property before it is a diagnosis: these are the parameters an operator would otherwise sweep per title, codec and operating point, and a system stable across their whole range ships with defaults instead ($\alpha=\beta=0.5$ throughout). The cause is structural --- Equations~\ref{eq:complexity}--\ref{eq:removability} blend terms that \emph{all} estimate how many bits a block costs, so they trade proxies for the same quantity.}

Grading the degradation \emph{strength} per block does not help either. Letting the score set strength as well as footprint costs bits on \num{7} of \num{8} videos at $b=\num{16}$, where Equation~\ref{eq:downsampling} admits only a single level, so we re-ran it where the bound permits three: $b=\num{64}$ at \num{1920}$\times$\num{1080}, sixteen sequences at four fixed QPs, differing in nothing but \texttt{downsample\_levels}. \textbf{A single strength is the better default: graded costs a median \qty{5.1}{\percent} of BD-rate on background LPIPS, losing on \num{14} of \num{16} at an exact two-tailed $p=\num{0.0042}$}, with foreground LPIPS differing by at most \num{0.0026}. The encoder pays for transitions between adjacent blocks held at different levels --- discontinuities the content never had. The design was sequential and is disclosed as such: pre-registered at $n=\num{10}$, returning \num{8} of \num{10} at $p=\num{0.109}$, then extended to a target of $n=\num{16}$ fixed before the extension ran, the six added sequences going \num{6} of \num{6} the same way. \textbf{What we retire is the graded \emph{default}, not the mechanism}: \texttt{tennis} ($\num{-11.1}\%$) and \texttt{motorbike} ($\num{-4.2}\%$) go the other way, and which content that is is the same unidentified property that decides the sign of reduction in Section~\ref{sec:results}. \texttt{camel} is excluded on a stability criterion rather than an outcome --- its graded arm drives more than a percent of output pixels to the clipping rails at every QP --- and that exclusion is itself a property of the graded transport (Appendix~\ref{sec:retired}).

\subsubsection*{Predicting post-restoration restorability}
\rev{An ideal rate-distortion-restoration rule would balance bits saved against post-restoration distortion, and the binding constraint is whether that second quantity is estimable \emph{before} transmission. The right panel of Figure~\ref{fig:selection-axes} answers it: across \num{120} restored runs on \num{13} sequences, ranking superblocks by spatial complexity ranks their eventual damage at a median Spearman $\rho = +\num{0.506}$, in the same direction on \num{120} of \num{120}. The score already selects on spatial complexity because complex blocks cost the most bits, so degrading them is the design working as intended. What the correlation adds is narrower: \textbf{within a fixed budget the ranking always spends it on the blocks that restore worst}. Over \emph{all} superblocks the correlation is \num{0.684}, because complexity predicts \emph{whether a block was touched} (\num{0.786}) before anything else; restricting to degraded blocks returns \num{0.506}, and that restriction was declared after the fact (Appendix~\ref{sec:provenance}).

We fit a damage predictor on five pre-registered transmit-time features, held out by video --- never by run, since runs of one clip share content and would leak --- reaching a held-out $\rho = +\num{0.400}$ on all \num{13} folds. \textbf{Restorability is therefore predictable at transmit time, with no client-side feedback.} To evaluate whether this feasibility translates directly into selection gains, we tested a first-order rate-distortion-restoration rule replacing the ranking with the ratio the diagnosis implies: bits freed over predicted damage. This changes about \qty{9}{\percent} of selected blocks and starts \qty{3.6}{\percent} down on rate; across six videos over four QPs, \textbf{it performs level with the heuristic incumbent} (median $+\num{5.1}\%$ BD-rate, better on \num{2} of \num{6}, $p=\num{0.69}$). This confirms the feasibility of transmit-time restorability modeling, establishes a baseline pricing for linear scoring, and indicates that realizing the remaining headroom requires joint non-linear optimization between rate cost and restorer distortion.}

\subsection{Spatial Selectivity vs.\ Uniform Downscaling}
\label{sec:downsample-vs-uniform}

Selective quality reduction with client-side reconstruction is not new: whole-frame downscaling with super-resolution at the client is deployed at scale. The architectural difference is that degradation here is per block under a mask excluding the foreground, giving the transport something uniform downscaling cannot express --- a region it will not touch. The uniform configuration is the identical code path with the budget opened to every block and protection off, so the per-block operator is pixel-identical; what differs is \emph{where} degradation is placed and \emph{how much} of the frame it covers, the uniform arm degrading roughly four times the area at a budget of \num{0.25}.

\rev{\textbf{Selectivity is ahead on two sequences and behind on four}, at $p=\num{0.69}$ on the foreground and $p=\num{0.22}$ on the background, \textbf{so we claim neither direction}; Appendix~\ref{sec:extra-tables} gives the per-sequence figures. On the four it loses it costs \qtyrange{60}{90}{\percent} more bits for equal perceptual quality, magnitudes the coverage difference predicts. \textbf{The comparison therefore establishes a lower bound on selectivity's standing:} it wins twice while giving away four times the degraded area.} Two limits bound it in opposite directions. The uniform arm sends full-resolution but low-pass-filtered frames rather than a genuine lower-resolution rendition with fewer coded pixels; a deployed ladder would likely do \emph{better}, so this is a conservative proxy for industry practice rather than a defeat of it. And the two sequences selectivity wins (\texttt{bear} and \texttt{india}) are exactly the two where uniform downscaling damages the foreground perceptibly ($+\num{0.087}$ and $+\num{0.101}$ foreground LPIPS) --- offered as a \emph{conjecture}, concordant on \num{6} of \num{6} but generated by these data and never pre-registered. Selectivity therefore behaves like insurance, costing rate where the foreground was never at risk and preventing a visible failure where it was. What the split does settle is negative: \textbf{foreground area does not predict which side a sequence falls on} --- \texttt{india} (\num{0.325}) wins and \texttt{pigs} (\num{0.327}) loses, \texttt{bear} (\num{0.114}) wins and \texttt{dog} (\num{0.090}) loses --- so the deciding quantity is how much perceptible damage a region takes and keeps.

\section{Conclusions}
\label{sec:conclusions}

Building on ELVIS, PRESLEY applies adaptive in-place degradation per block under a removability mask, signals per-block strength in a bit-packed side channel, and restores with a generative model conditioned on transmitted visual priors rather than unconditioned in-painting. Of the degradation modalities we priced, progressive downsampling survives; QP mapping and noise injection are retired with the empirical measurements that rule them out.

Read against the three goals, the evaluation demonstrates clear, definitive advancements alongside precise domain boundaries. On \emph{restoration}, conditioned super-resolution delivers the best background quality of any configuration we tested, beating hole-filling in-painting on every video, with foreground fidelity held bit-exact by passthrough compositing. On \emph{reduction}, PRESLEY surpasses its predecessor at matched rate on \num{13} of \num{13} ladders, achieving a mean \qty{-56.4}{\percent} BD-rate and demonstrating substantial coding gains against standard codecs in the target bit-starved regime (\num{17} of \num{23} clips). On \emph{selection}, our oracle analysis bounds the cost axis at ${\approx}\qty{5}{\percent}$ remaining headroom and establishes transmit-time damage predictability as the definitive frontier for future generative streaming architectures. While a first-order linear ratio performs level with the incumbent, it proves the feasibility of the concept and defines joint non-linear optimization as the next step.

\rev{Two limitations bound all of this. Restoration is far from real time (\numrange{6.3}{61.5}$\times$ slower than the source) and the cheapest optimisation is already taken, the restorer running in half precision by default, so we scope the system to offline transcoding and edge-cache pre-population. And every perceptual claim rests on learned metrics rather than human raters: no subjective study was conducted.}

Three directions follow from that map. The first is a restorer built \emph{for} this transport rather than adapted to it: every backbone we ran was trained for whole-frame super-resolution or in-painting, then handed a frame degraded per block, at a strength the side channel already carries, in rounds. A model consuming that strength map natively --- restoring block by block, once per level, conditioned on how far each block was taken and on its undegraded neighbours --- would use information the pipeline already transmits and no current backbone reads. The second is the content property deciding whether reduction pays, since foreground area is refuted and identifying it would do more here than another codec or backbone. The third is a selection rule spending its budget on the restorability term we measure, which is predictable at transmit time and sits at the cheapest point in the pipeline to change. Two smaller openings: the protection mask is a segmenter's notion of what matters, where a viewer-specific one would fit the same interface and address the personalization barrier that has held back region-of-interest coding; and a subjective study is the natural way to confirm that what our metrics reward is what a viewer sees.

\begin{acks}
The financial support of the Austrian Federal Ministry for Digital and Economic Affairs, the National Foundation for Research, Technology and Development, and the Christian Doppler Research Association, is gratefully acknowledged.
Christian Doppler Laboratory ATHENA: \url{https://athena.itec.aau.at/}
\end{acks}

\printbibliography

\clearpage
\section{Appendix}

\appendix

\section{Evaluation Principles and Methodological Pitfalls}
\label{sec:pitfalls}

We document five key methodological principles and pitfalls established during system evaluation to assist researchers developing and validating generative video streaming pipelines:

\subsection*{1. Region metrics inadvertently evaluating background}
Evaluating foreground DISTS on a union bounding box (averaging \qty{76}{\percent} of frame area vs.\ \qty{15}{\percent} true foreground) scored the background that the transport intentionally degraded. Correcting to mask-weighted spatial feature pooling makes our own foreground scores \emph{worse} on \num{30} of \num{35} sequences, and changes the winner in \num{34} matched-rate groups --- \num{16} toward the degrade-and-restore configurations and \num{12} toward the baseline. \textbf{Principle:} Always verify the true spatial coverage of region metrics; bounding boxes do not constitute isolated regions.

\subsection*{2. Variable bitrate obscuring bitrate reductions}
Under VBR rate control, the encoder expends its target bitrate regardless of source complexity, preventing block degradation from freeing bandwidth and instead causing degradation artifacts to steal bits from the foreground. Across every matched pair in our corpus, VBR degradation costs \qty{16.4}{\percent} more bits than the pristine baseline, saving on \num{0} of \num{18}, where fixed QP frees \qty{9.2}{\percent}. \textbf{Principle:} Fixed-QP rate control is mandatory for evaluating degradation transports. A degradation curve measured under a bitrate target measures the rate controller, not the method. We enforce this in code --- any VBR degradation run is flagged as uncitable automatically.

\subsection*{3. One-tailed significance testing bias}
Early analyses quoted one-tailed $p$-values based on post-hoc observations of empirical directionality. \textbf{Principle:} Two-tailed always, and know the floor --- at $n=5$ no split can reach \num{0.05}, so a consistent small suite is underpowered rather than negative.

\subsection*{4. Device population mixing in latency benchmarks}
A silent CPU fallback produced artificial \numrange{10}{40}$\times$ latency spreads by mixing CPU and GPU executions. \textbf{Principle:} Explicitly log hardware device backends per trial and disallow heterogeneous averaging.

\subsection*{5. Inconsistent rate control comparisons}
Comparing fixed-QP ROI configurations against VBR baselines misattributed baseline overshooting as ROI savings. \textbf{Principle:} Compare configurations only against like-for-like rate control baselines.

\section{Metrics and Statistical Protocol}
\label{sec:metrics}

\subsection*{Spatial mask restriction for deep perceptual metrics}
For LPIPS and DISTS, per-frame foreground masks are resampled onto the network's spatial feature maps and utilized as location-wise weights in internal pooling, assigning zero weight to background coordinates. Setting all weights to one identically reproduces standard whole-frame metrics. Because deep convolutional receptive fields span tens of pixels, mask-weighted pooling evaluates foreground objects within their local scene context without introducing boundary cropping artifacts. For the same reason no foreground claim rests on VMAF or FVMD, whose available foreground variants are bounding-box crops. Background PSNR is reported but is never a verdict: the flat mean fill scores the best background PSNR of any fill we tested while being the worst-looking, because a smooth patch is arithmetically closer to the original than invented texture is.

\subsection*{Statistical significance protocol}
\rev{No subjective study was conducted: every perceptual claim in this article rests on mask-restricted learned metrics, and none of them substitutes for human raters. We also do \emph{not} gate claims on a perceptual threshold. We are aware of no source establishing \num{0.05} as a calibrated just-noticeable difference for LPIPS or DISTS, which are in any case on different scales, so wherever a \num{0.05} margin appears in this article it is a declared reporting margin --- applied identically to our configurations and to every baseline --- and never a perceptual constant.}

Effect sizes are therefore reported directly, and claims are gated on paired significance instead: an exact two-tailed sign test over videos, Holm-corrected across every candidate this project tried, losers included, with the number of pairs stated. Where a comparison cannot reach significance --- at six paired videos the smallest attainable two-tailed $p$ is \num{0.031}, and below six no split reaches \num{0.05} --- we report it as underpowered rather than as a null.

\section{Super-Resolution Restoration Algorithm}
\label{sec:restoration-algs}

Algorithm~\ref{alg:sr_round} details the progressive super-resolution restoration procedure on the client. Transmitted pixels are re-injected for blocks that have reached their target resolution, preserving un-degraded content and ensuring foreground quality remains independent of the restorer backbone.

\begin{algorithm}[h]
\caption{Progressive Super-Resolution Restoration}
\label{alg:sr_round}
\small
\begin{algorithmic}[1]
\Require $F$: current frame at $1/2s$ resolution; $F_{in}$: decoded frame; $D \in \mathbb{Z}^{I \times J}$: downsample factors; $s$: current scale
\Ensure $F_{up}$, $D$: upscaled frame and updated factors
\State $F_{up} \gets \text{Model}(F)$ \Comment{Conditioned $2\times$ SR pass}
\State $F_{ref} \gets \text{Resize}(F_{in}, \text{size}(F_{up}))$
\For{$i = 0$ \textbf{to} $I - 1$}
    \For{$j = 0$ \textbf{to} $J - 1$}
        \If{$D_{i,j} \le s$}
            \State $F_{up}[i, j] \gets F_{ref}[i, j]$ \Comment{Preserve transmitted pixels}
        \Else
            \State $D_{i,j} \gets s$ \Comment{Update scale factor}
        \EndIf
    \EndFor
\EndFor
\State \Return $F_{up}$, $D$
\end{algorithmic}
\end{algorithm}

\section{Architectural Design Principles and Retired Candidates}
\label{sec:retired}
\label{sec:extra-tables}

\subsection*{Architectural principles for neural video streaming transports}
Each of the following candidate modalities was evaluated and retired based on rigorous empirical measurements. Together, they establish foundational architectural design principles for neural video streaming transports, preventing future wasted research effort:
\begin{enumerate}
    \item \textbf{Avoid frame shrinking (preserve spatial geometry):} Repacking the most removable blocks out of the frame appears to offer a bitrate win by transmitting roughly \qty{25}{\percent} fewer pixels, but instead costs \qtyrange{42.8}{47.7}{\percent} \emph{more} bits than in-place blackout at the same budget. Repacking destroys intra- and inter-frame spatial prediction. In-place degradation is essential.
    \item \rev{\textbf{Degradations must align with transform coding (avoid noise injection):} Adding high-frequency Gaussian noise costs \qtyrange{76.5}{83.0}{\percent} more bits than the pristine baseline at indistinguishable foreground quality, whereas same-budget downsampling and blur \emph{free} \qtyrange{10}{27}{\percent}. Degradation must reduce entropy for the codec, not merely appear smooth to human vision; noise severely disrupts motion compensation.}
    \item \rev{\textbf{ROI control belongs in the encoder core (avoid pixel-plane QP mapping):} Setting per-block QP offsets from removability re-quantizes coefficients on the pixel plane \emph{before} encoding rather than driving the encoder's native quantization engine. Combined with GoP averaging, it yields inferior rate-distortion trade-offs compared to native codec ROI (Kvazaar gains $+0.42$~dB FG at neutral rate; Appendix~\ref{sec:roi-detail}). In-place pixel downsampling is the superior pre-encode transport.}
    \item \textbf{Prioritize low-entropy fills for in-painting transports:} Foreground PSNR is flat across blackout, freeze, and mean-fill on \num{6}/\num{6} videos due to passthrough compositing, and background LPIPS does not separate them. The choice is settled on rate: flat blackout is cheapest at \qty{566.9}{kbps} against \qty{589.3}{} and \qty{616.1}{}.
    \item \rev{\textbf{Hybrid codecs remain essential (neural video representations like HNeRV are non-viable for streaming):} Accounting for transmitted serialized model weights reveals that HNeRV requires \numrange{22.7}{178.1}$\times$ higher bitrate than standard x265 at matched foreground quality, confirming that hybrid traditional codecs coupled with lightweight client-side generative layers are the practical path forward.}
    \item \textbf{Enforce fixed-QP rate control:} Fixed-QP is the prerequisite for bit relocation; under VBR, the rate controller launders bits regardless of degradation.
\end{enumerate}

One sequence is excluded from the graded-strength comparison of Section~\ref{sec:ablation} on a stability criterion rather than an outcome, and the exclusion is itself evidence about the graded transport. \texttt{camel}'s graded configuration drives more than a percent of its output pixels to the clipping rails at every QP, because a multi-level pyramid applies the restorer once per level and each round re-amplifies what the last pushed toward the rails. It breaches the clipping invariant on all four rungs, against zero breaches across the \num{40} binary runs.

\subsection*{Progression across pipeline stages}
Table~\ref{tab:chain} reports the stage-by-stage progression from native codec ROI to ELVIS block removal and PRESLEY degrade-and-restore. Table~\ref{tab:downsample-vs-uniform} presents the per-sequence performance comparison between spatially selective and whole-frame uniform downsampling.

\begin{table*}[t]
\centering

\begin{minipage}[t]{0.48\textwidth}\centering
\caption{Stage-by-stage progression vs.\ pristine fixed-QP baselines at \num{640}$\times$\num{360} (median over 6 sequences).}
\label{tab:chain}
\small
\begin{tabular}{llrrr}
\toprule
Stage & Encoder & $\Delta$bits & $\Delta$FG & $\Delta$BG \\
\midrule
Codec ROI & kvazaar & $-1.0\%$  & $+0.42$~dB & $-0.74$~dB \\
ELVIS & SVT-AV1 & $+22.8\%$ & $-0.15$~dB & $-5.47$~dB \\
PRESLEY & SVT-AV1 & $+1.2\%$ & $-0.13$~dB & $-1.85$~dB \\
\midrule
ELVIS & x265 & $-1.7\%$ & $-0.16$~dB & $-9.43$~dB \\
\bottomrule
\end{tabular}
\end{minipage}\hfill
\begin{minipage}[t]{0.48\textwidth}\centering
\caption{Selective vs.\ uniform downsampling (SVT-AV1, \num{640}$\times$\num{360}, bs8, Real-ESRGAN). Negative BD-rate favors selective.}
\label{tab:downsample-vs-uniform}
\small
\setlength{\tabcolsep}{3pt}
\begin{tabular}{lrrrr}
\toprule
Video & BD-rate FG & BD-rate BG & $\Delta$FG-LPIPS & Overlap \\
\midrule
\texttt{bear}   & $-18.3\%$ & $-10.4\%$ & $+0.087$ & 0.83 \\
\texttt{india}  & $-30.8\%$ & $+40.5\%$ & $+0.101$ & 0.67 \\
\midrule
\texttt{camel}  & $+60.2\%$ & $+21.7\%$ & $-0.001$ & 0.67 \\
\texttt{tennis} & $+60.6\%$ & $+65.6\%$ & $-0.027$ & 0.65 \\
\texttt{pigs}   & $+86.4\%$ & $+51.8\%$ & $-0.022$ & 0.60 \\
\texttt{dog}    & $+89.8\%$ & $+55.5\%$ & $-0.032$ & 0.62 \\
\bottomrule
\end{tabular}
\end{minipage}
\end{table*}

\section{Robustness Checks: Segmentation and Resolution}
\label{sec:robustness}

\begin{table}[t]
\centering
\caption{Detector-derived mask (YOLOE-11l-seg) vs.\ ground truth on 7 sequences.}
\label{tab:mask-area}
\small
\begin{tabular}{lrrr}
\toprule
Sequence & Area ratio & $\Delta$bitrate & $\Delta$FG-PSNR \\
\midrule
\texttt{india}       & $2.65$ & $+16.4\%$ & $+2.82$~dB \\
\texttt{camel}       & $1.09$ & $-2.3\%$  & $+0.47$~dB \\
\texttt{bear}        & $0.91$ & $-2.2\%$  & $-0.12$~dB \\
\texttt{pigs}        & $0.88$ & $-3.8\%$  & $-0.36$~dB \\
\texttt{bmx-trees}   & $0.86$ & $-10.1\%$ & $-2.90$~dB \\
\texttt{tennis}      & $0.86$ & $-8.0\%$  & $-4.37$~dB \\
\texttt{dog}         & $0.54$ & $-0.4\%$  & $-0.66$~dB \\
\bottomrule
\end{tabular}
\end{table}

\begin{figure*}[t]
    \centering
    \includegraphics[width=0.92\textwidth]{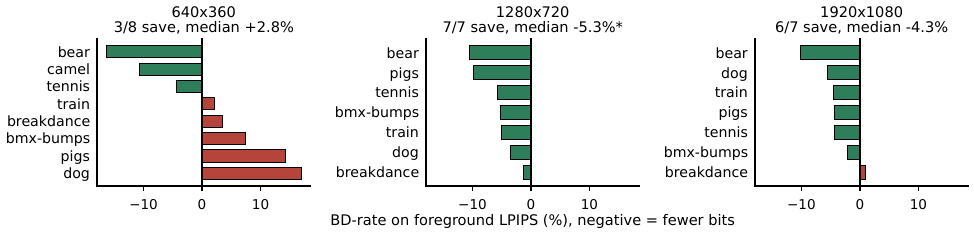}
    \input{Figures/resolution_ladder.desc.tex}
    \caption{BD-rate on foreground LPIPS per clip across resolutions (\num{80}$\times$\num{45} grid).}
    \label{fig:resolution-ladder}
\end{figure*}

\subsection*{Detector-derived foreground masks}
\rev{Replacing DAVIS ground-truth masks with detector masks (YOLOE-11l-seg, Table~\ref{tab:mask-area}) indicates that foreground quality follows the area a mask protects, not the accuracy with which it segments: the rank correlation is $\rho = +0.893$, and \texttt{india} --- the one sequence whose detector mask is larger than the annotation --- is also the one whose bitrate rises. Morphological perturbations (dilation/erosion radius 4) alter foreground PSNR by $\le \num{0.12}$~dB, demonstrating robustness to boundary noise.}

\subsection*{Resolution scaling}
\textbf{At \num{1280}$\times$\num{720} the saving is a result rather than a trend}: all \num{7} citable clips need fewer bits for equal foreground quality, a median of \qty{-5.3}{\percent}, exact two-tailed $p=\num{0.0156}$, or \num{0.047} after Holm correction across the three resolutions. \textbf{At \num{1920}$\times$\num{1080} it does not follow} (\num{6} of \num{7}, median \qty{-4.3}{\percent}, $p=\num{0.125}$), and at \num{640}$\times$\num{360} the split persists at \num{3} of \num{8}. Two things bound this. At fixed QP the higher resolutions run at roughly half the bits per pixel and so sit in the \emph{more} starved regime that favours the method; and \texttt{breakdance} spends \num{0.214} bits per pixel at \num{640}$\times$\num{360} against the \numrange{0.078}{0.136} of the other seven, outside the range registered in advance as one operating regime, so its lowest-quality QP is reported but not pooled.

\section{Whole-Frame Quality Metrics}
\label{sec:wholeframe}

Table~\ref{tab:wholeframe} reports full-frame metric changes against pristine baselines across seven standards. Learned metrics (LPIPS, DISTS) and structural metrics confirm the ranking established by region-restricted evaluations.

\begin{table*}[t]
\centering
\caption{Whole-frame metrics: median change against pristine baselines paired within (video, codec, QP, resolution).}
\label{tab:wholeframe}
\small
\begin{tabular}{lrrrrrrrr}
\toprule
Configuration & $n$ & $\Delta$PSNR$\uparrow$ & $\Delta$SSIM$\uparrow$ & $\Delta$VMAF$\uparrow$ & $\Delta$LPIPS$\downarrow$ & $\Delta$DISTS$\downarrow$ & $\Delta$FID$\downarrow$ & $\Delta$FVMD$\downarrow$ \\
\midrule
PRESLEY downsample + Real-ESRGAN & \num{295} & $-2.21$  & $-0.053$ & $-7.8$  & $+0.026$ & $+0.003$ & ---     & ---     \\
PRESLEY blackout + ProPainter    & \num{18}  & $-7.18$  & $-0.117$ & $-36.3$ & $+0.129$ & $+0.047$ & $+12.2$ & $+1368$ \\
PRESLEY freeze + ProPainter      & \num{28}  & $-9.50$  & $-0.200$ & $-51.4$ & $+0.156$ & $+0.052$ & $+12.8$ & $+2772$ \\
ELVIS (block removal)            & \num{149} & $-11.67$ & $-0.195$ & $-45.2$ & $+0.177$ & $+0.070$ & $+16.2$ & $+2284$ \\
\bottomrule
\end{tabular}
\end{table*}

\section{Per-Point Codec ROI Performance}
\label{sec:roi-detail}

Table~\ref{tab:roi} details every evaluated operating point for Kvazaar native ROI encoding.

\begin{table}[h]
\centering
\caption{Kvazaar native ROI encoding per operating point under fixed QP.}
\label{tab:roi}
\scriptsize
\setlength{\tabcolsep}{3.5pt}
\begin{tabular}{llrrrrr}
\toprule
Video & Setting & Rate ROI & Rate base & $\Delta$Rate & $\Delta$FG-PSNR & $\Delta$BG-PSNR \\
\midrule
bear          & kv-150   & 155.8 & 146.5 & $+6.3\%$  & $+0.53$ dB & $-0.40$ dB \\
bear          & kv-250   & 233.7 & 224.5 & $+4.1\%$  & $+0.57$ dB & $-0.53$ dB \\
bear          & kv-460.8 & 460.5 & 492.4 & $-6.5\%$  & $+0.71$ dB & $-0.73$ dB \\
bear          & kv-555   & 577.5 & 492.4 & $+17.3\%$ & $+1.34$ dB & $-0.22$ dB \\
bear          & kv-800   & 727.1 & 823.9 & $-11.7\%$ & $+0.73$ dB & $-0.89$ dB \\
bmx-trees     & kv-460.8 & 501.1 & 467.0 & $+7.3\%$  & $+0.24$ dB & $-0.04$ dB \\
camel         & kv-300   & 304.0 & 279.8 & $+8.7\%$  & $+0.51$ dB & $-1.12$ dB \\
camel         & kv-600   & 572.4 & 588.1 & $-2.7\%$  & $+0.58$ dB & $-1.28$ dB \\
dog           & kv-300   & 282.4 & 328.9 & $-14.1\%$ & $+0.57$ dB & $-0.88$ dB \\
dog           & kv-600   & 593.0 & 616.5 & $-3.8\%$  & $+1.00$ dB & $-0.62$ dB \\
drift-chicane & kv-460.8 & 447.1 & 447.4 & $-0.1\%$  & $+0.21$ dB & $-0.13$ dB \\
india         & kv-300   & 299.3 & 290.8 & $+2.9\%$  & $+0.23$ dB & $-0.20$ dB \\
india         & kv-600   & 577.1 & 565.6 & $+2.0\%$  & $+0.20$ dB & $-0.25$ dB \\
pigs          & kv-300   & 311.2 & 322.6 & $-3.5\%$  & $+0.13$ dB & $-0.95$ dB \\
pigs          & kv-600   & 553.3 & 654.4 & $-15.4\%$ & $+0.09$ dB & $-1.33$ dB \\
tennis        & kv-300   & 300.1 & 315.5 & $-4.9\%$  & $-0.45$ dB & $-0.25$ dB \\
tennis        & kv-600   & 622.1 & 581.7 & $+6.9\%$  & $+0.14$ dB & $+0.13$ dB \\
\bottomrule
\end{tabular}
\end{table}

\section{Data Provenance and Pre-registration Commitments}
\label{sec:provenance}

Every experiment record in this article is tracked by an immutable configuration hash and validated against five automated invariants: existence of valid region metrics, bitrate accounting within \qty{1}{\percent}, fixed-QP rate control, non-degraded background restoration, and absence of numerical saturation. Runs failing any invariant are excluded.

Three analyses were additionally bounded before the runs they depend on existed, with the design documents committed ahead of the data, and \textbf{we record what fired rather than revising the bounds}. The regime-stability contrast did not fire ($p=\num{0.58}$), which is the reported result. For restorability, the correlation over \emph{all} superblocks exceeded its band, and the control that resolved it --- restricting to actually-degraded blocks, which is where the reported $\rho = +\num{0.506}$ comes from --- was declared after the fact, so both figures are reported. For the resolution ladder, two sequences fell outside the band on foreground BD-rate and one further bound fired, on \texttt{breakdance}'s bits per pixel. Two further bounds fired elsewhere in the article and are disclosed where their results appear: the placement contrast of Section~\ref{sec:exclusion} landed far outside a band that was itself mis-derived, and that section's exclusion test was pre-registered on foreground LPIPS while its stated basis was a foreground PSNR figure. Both mis-specifications are ours.

\end{document}